\documentclass[times,twocolumn]{aastex63}
\usepackage[utf8]{inputenc}
\usepackage[normalem]{ulem}
\usepackage{amsmath}
\usepackage{array,multirow,graphicx}
\usepackage{verbatim}

\newcommand{\HI}{H\,{\sc i}\ }
\newcommand{\HII}{H\,{\sc ii}\ }
\newcommand{\model}{NE2025}
\newcommand{\DM}{{\rm DM}}
\newcommand{\SM}{{\rm SM}}
\newcommand{\nelec}{n_{\rm e}}
\newcommand{\Fc}{F_{\rm c}}
\newcommand{\Nxgal}{N_{\rm > mod}}       

\newcommand{\be}{\begin{eqnarray}}
\newcommand{\ee}{\end{eqnarray}}

\newcommand{\thetad}{\theta_{\rm d}}

\newcommand{\dsl}{d_{\rm sl}}

\newcommand{\xtau}{x_{\tau}}
\newcommand{\revision}{\textcolor{black}}

\begin{document}

\title{NE2025: An Updated Electron Density Model for the Galactic Interstellar Medium}

\author[0000-0002-4941-5333]{Stella Koch Ocker}
\email{socker@caltech.edu}
\affiliation{Cahill Center for Astronomy and Astrophysics, California Institute of Technology, Pasadena, CA 91125, USA}
\affiliation{Observatories of the Carnegie Institution for Science, Pasadena, CA 91101, USA}

\author[0000-0002-4049-1882]{James M. Cordes}
\affiliation{Department of Astronomy and Cornell Center for Astrophysics and Planetary Science, Cornell University, Ithaca, NY, 14853, USA}

\correspondingauthor{Stella Koch Ocker}

\begin{abstract} 
    Free electrons in the Galactic interstellar medium (ISM) disperse and scatter coherent radio waves, by amounts that depend on the distance to the radio source. Models of the Galactic electron density are thus widely used to predict distances and scattering of compact radio sources (including pulsars, fast radio bursts (FRBs), and long-period transients), in addition to mitigating ISM foregrounds in Galactic and extragalactic studies. We use a sample of 171 precise pulsar distances, based entirely on parallaxes and globular cluster associations, as well as scattering measurements of 568 pulsars, active galactic nuclei, and masers, to update the NE2001 Galactic electron density model. We refit the thick and thin disks and three of the spiral arms. The new parameters for these large-scale components significantly repartition free electrons between the thick disk and spiral arms, thereby correcting NE2001's systematic underestimation of pulsar distance and scattering. Sightlines with excessive dispersion and scattering are used to identify new clumps that are added to the model, in addition to refining clumps that were already included (e.g., Cygnus, Vela, and Gum). The Galactic Center component is revised, yielding scattering time predictions that are $10^3$ times smaller than the Galactic Center in NE2001. The updated model, \model, provides a factor of $20\times$ improvement in median distance prediction accuracy and $100\%$ median improvement in scattering predictions based on \revision{dispersion measure}, relative to NE2001. There is a $15\times$ improvement in median distance prediction accuracy relative to YMW16. \model\ is available on Github and the Python Package Interface. 
\end{abstract}

\keywords{Interstellar medium -- Warm ionized medium -- Milky Way Galaxy -- Galactic Center -- Pulsars -- Radio bursts -- Radio active galactic nuclei -- Magnetars -- Hydroxyl masers}

\section{Introduction}

Galactic electron density models describe the ionized gas distribution of the Milky Way interstellar medium (ISM). Existing models have primarily been calibrated using the dispersion measure (or electron column density; ${\rm DM} = \int n_e dl$), scattering, and distances of compact radio sources, including pulsars and active galactic nuclei (AGN). Modeling of the ISM electron density on a Galaxy-wide basis has been an ongoing activity since the discovery of pulsars nearly 60 years ago \citep[][]{hewish1968}. Initially, only the local mean density was estimated and then models with smooth, large scale disk structures were progressively developed.  These, in turn,  were augmented with large and small scale features as demanded by measurements along individual lines of sight (LOSs).  \citet[][]{prentice-terhaar1969} recognized early on that the LOSs to some pulsars traversed local \HII regions but these were not incorporated into Galactic models until later. Early models incorporated azimuthally symmetric disks, typically with exponential dependences in $\vert z \vert$ transverse to the Galactic plane and a power-law or Gaussian dependence in Galactocentric radius $R$ \citep[e.g.][]{1977ApJ...215..885T, 1981AJ.....86.1953M}. \citet[][]{1982ApJ...257..603H} incorporated thin and thick disk components having different radial dependences and offsets from $z=0$ as a function of $R$. Though termed a `warp,' this structure was fully azimuthally symmetric, unlike the Milky Way warp identified more recently (see \citealt{kalberla09} for a review).  The thick component also had a smaller $z$ scale height in the inner Galaxy compared to near the solar circle and the thin disk was annular in form with no contribution inside $R = 4$\,kpc. A simpler model was presented by \citet{1982JApA....3..399V} as a constant density (corresponding effectively to a thick disk with indefinite scale height) and a thin plane-parallel component. \citet[][]{1985MNRAS.213..613L} added a strong feature associated with the Gum Nebula to an otherwise azimuthally symmetric model of the same form as used in
\citet[][]{1981AJ.....86.1953M}. Spiral arms accompanied axisymmetric components previously incorporated by other authors in the model of \citet[][TC93]{1993ApJ...411..674T}.  Another innovation of the TC93 model, based on work in \citet[][]{cordes91}, was the inclusion of plasma scattering observables in model fitting along with dispersion and distance data. The discovery of a large number of pulsars in the  Parkes multibeam pulsar survey \citep[][]{manchester2001} stimulated further model development, albeit with reversion to an axisymmetric model by
\citet[][]{gomez2001}.

One of the most widely used Galactic electron density models is NE2001 \citep{ne20011,ne20012}. This comprehensive model built upon the basic structure of TC93 but added several new features: a local ISM component, a Galactic Center component, a large number of clumps for LOSs with known \HII regions, supernova remnants, and for directions with manifested enhanced scattering. Voids were also introduced to account for the distances of some objects requiring an underdensity  or a low scattering level. Kolmogorov turbulence was explicitly included in order to forward model radio wave scattering observables (see \citealt{ne20011} for details). NE2001 and other Galactic models, including the more recent YMW16 \citep{ymw16}, have been used  routinely for estimating pulsar distances, for forecasting the effects of scattering in surveys for pulsars, fast radio bursts (FRBs), and extraterrestrial intelligence, and for calculating the Galactic contribution to the DM inventories of FRBs.   While not explicitly designed for ISM studies, NE2001 has been used in model construction for Galactic gamma-ray emission and for contributions to the Galactic foreground in studies of the cosmic microwave background. 

All of the electron density models described above performed model-fitting with limited samples of independent pulsar distance measurements, the majority of which were not based on precise ($<25\%$ fractional uncertainty) parallaxes.\footnote{Of the 112 distances used to fit NE2001, just 14 were based on parallax; the majority of distances used were based on \HI kinematics.} The sample of precise pulsar parallaxes has since grown substantially \citep{2009ApJ...698..250C,deller2019,ding2023}, and is compounded by an increase in precise distances based on pulsar associations with globular clusters \citep{pan2021_globs,ridolfi2022,padmanabh2024}. These new, high-precision distance samples have revealed key areas for improvement in Galactic electron density models, including systematic underestimation of distances for sightlines at high Galactic latitudes \citep{price2021}. Concurrently, comparison between the pulsar population and cataloged \HII regions has revealed that as much as one-third of known pulsar sightlines may intersect \HII regions, and that \HII regions account for nearly all Galactic sightlines with DMs that exceed the maximum predictions of NE2001 and YMW16 \citep{ocker2024c}.

\begin{figure*}
    \centering
    \includegraphics[width=0.85\textwidth]{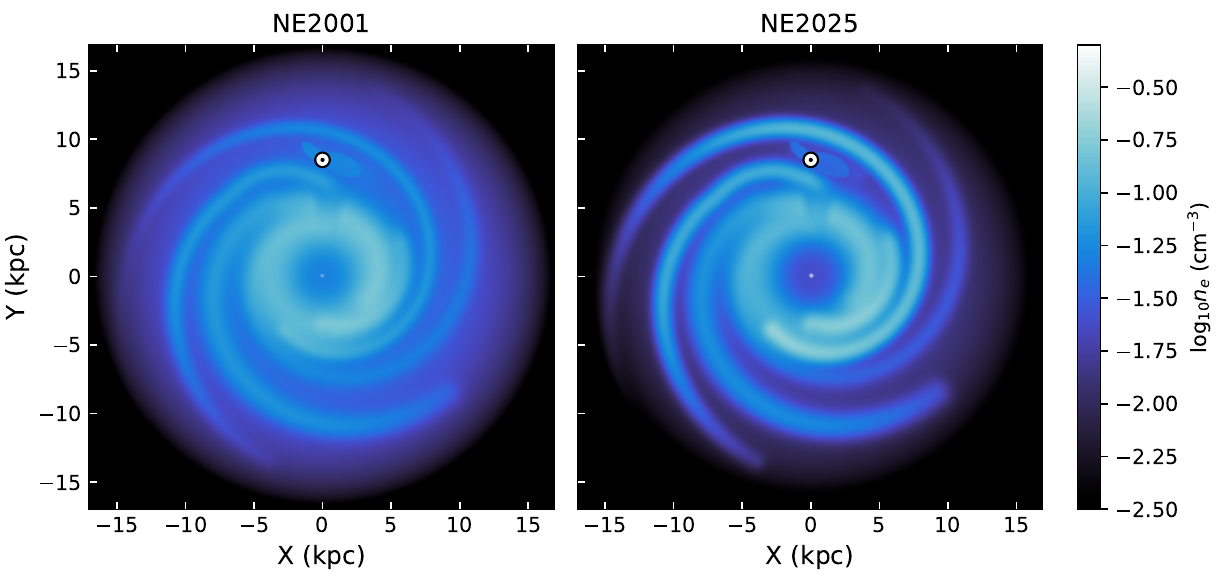}
    \caption{Electron density in the Galactic plane ($Z = 0$) for NE2001 (left) and \model\ (right). The color stretch is the same for both panels.}
    \label{fig:electron-map}
\end{figure*}

These developments motivate a reassessment of the NE2001 model parameters. This study delivers an updated model, \model, which uses the original NE2001 framework but with revised parameters and extensions to specific model components. Figure~\ref{fig:electron-map} presents a comparison of the electron density in the Galactic plane between the NE2001 and \model\ models. Among the notable improvements in \model\ are revisions to the thick and thin disks and multiple spiral arms, resulting in substantial reductions in the number of pulsars for which the model predicts observables that differ from actual measurements. Unique to both NE2001 and \model, in comparison to other models, is the explicit treatment of scattering due to electron density fluctuations that accompany local mean densities across the Galaxy. \model\ is provided through the same package distribution as NE2001p, the Python implementation of NE2001 \citep{ocker2024_ne2001p}, as well as in Fortran. All parameter values quoted in this paper are evaluated using the Fortran version of each model. The updated model presented here provides near-term improvement in accuracy compared to NE2001, in anticipation of pulsar astrometry programs and ISM surveys that will deliver expanded datasets informing a next-generation model \revision{already in progress}.

The layout of the paper is as follows: Section~\ref{sec:components} briefly summarizes the model architecture. In Section~\ref{sec:model-comparison}, we describe the pulsar distance sample and compare NE2001 predictions to the observed distributions of pulsar distance, DM, and scattering. These comparisons motivate revisions to the model based on fitting criteria described in Section~\ref{sec:fitting}. We then elaborate on the refitted model components, including the thick disk (Section~\ref{sec:thick}), thin disk and spiral arms (Section~\ref{sec:spiral}), the Gum Nebula and Vela supernova remnant (Section~\ref{sec:local}), the Galactic Center (Section~\ref{sec:GC}), and new clumps and voids (Section~\ref{sec:clumps}). Conclusions are discussed in Section~\ref{sec:conc}.

\section{Model Components}\label{sec:components}

The structural components of \model\ are identical to those in NE2001 as summarized in Table\,2 of \citet[][]{ne20011}. However we present significant changes in parameter values and also alter many of the small-scale features in the model.  Briefly, the large scale components comprise thin and thick disk components along with five spiral arms.  Prominent individual regions represent  the local ISM and the volume surrounding the Galactic Center. Small regions include `clumps'  with enhanced electron densities (most likely due to \HII\ regions and supernova remnants) primarily in the thin disk and spiral arms. Underdense regions (`voids') required along some LOSs  are likely related to supernova driven bubbles and chimneys. 

Essentially all model elements have been revisited, with major changes to the thick disk and Galactic Center, and smaller changes to the thin disk and spiral components.  The rosters of clumps and voids have been altered with addition of prominent clumps along the direction to the Galactic Center and to the Cygnus region.   The general region of the Vela supernova remnant and Gum Nebula has been altered significantly.    Finally, some clumps and voids have been removed and others added, some as replacements. 

\section{Advancing  the Model with New Pulsar Observations}\label{sec:model-comparison}

\begin{figure}
    \centering
    \includegraphics[width=0.45\textwidth]{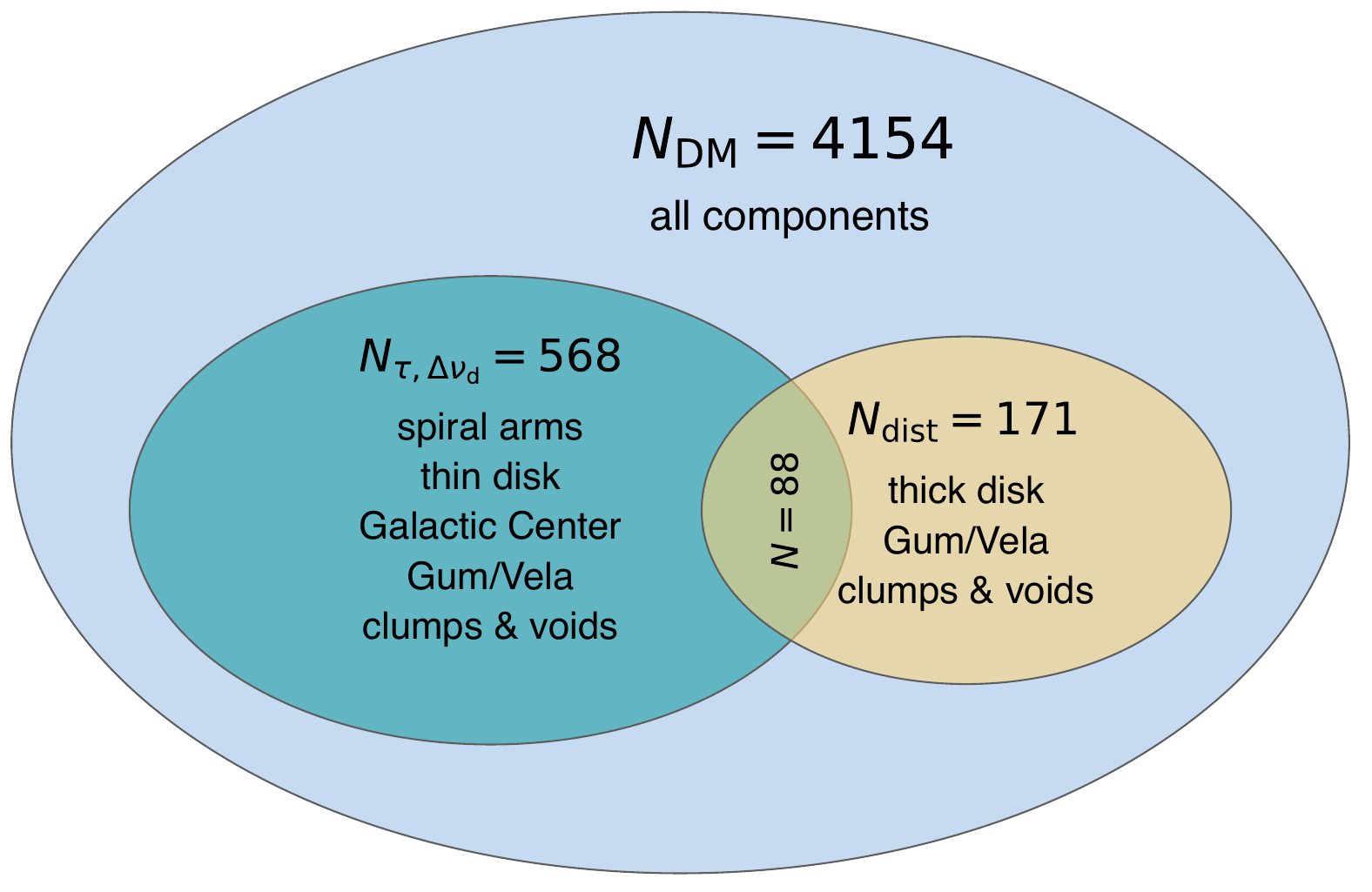}
    \caption{\revision{The primary datasets used to fit \model. Numbers of DM, scattering, and distance measurements are indicated, as well as the model components constrained by each dataset. See Section~\ref{sec:model-comparison} for details.}}
    \label{fig:venn}
\end{figure}

\revision{The primary datasets used to refit NE2001 are (1) the sample of independent pulsar distances, (2) the DM distribution of the entire known Galactic radio pulsar population, constituting $\approx 4200$ radio pulsars in v2.7 of the ATNF Pulsar Catalog \citep{psrcat}, and (3) scattering measurements for 568 pulsars that we have compiled from the literature. Figure~\ref{fig:venn} shows a Venn diagram of the overlap between these datasets, and the model components they constrain. In the following sections, we further describe these datasets and use them to evaluate the NE2001 model performance, leading to the identification of specific model components that are significantly discrepant from the data, and motivating the fitting procedure discussed in Section~\ref{sec:fitting}. In subsequent sections that treat high-density regions, namely the Galactic Center (Section~\ref{sec:GC}) and Cygnus (Section~\ref{sec:cygnus}), we additionally assess scattering measurements of masers, AGN, and FRBs.}

\subsection{Independent Distance Sample}\label{sec:distance-sample}

\begin{figure*}
    \centering
    \includegraphics[width=\textwidth]{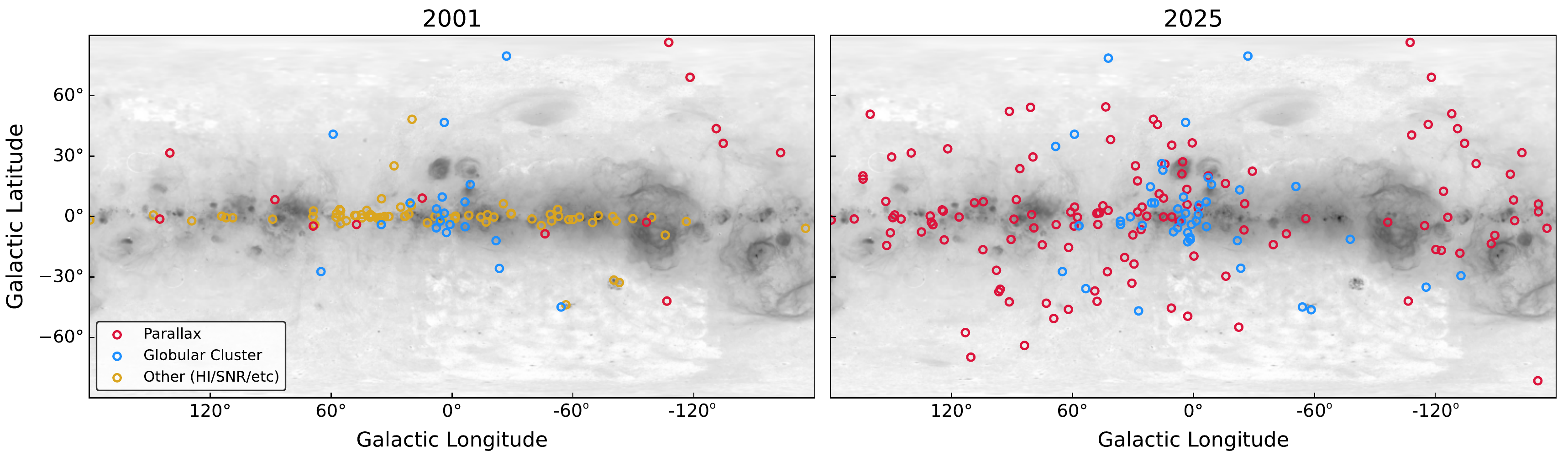}
    \caption{Comparison between the sky distributions of distance measurements used to calibrate NE2001 and \model. Left: Pulsar parallaxes (red), globular cluster associations (blue), and other distance measures (yellow; e.g., \HI kinematic distances and supernova remnant (SNR) associations) used to calibrate NE2001 are shown on top of an all-sky H$\alpha$ intensity map (grey) from \cite{2003ApJS..146..407F}. Right: All parallaxes and globular cluster associations used to update NE2001.}
    \label{fig:2001-2025-comparison}
\end{figure*}

\revision{One of the datasets} we use to evaluate and refit NE2001 is the sample of independent pulsar distances. We use two of the most precise types of pulsar distance measurement: parallaxes and globular cluster associations. While pulsar distance measures based on \HI kinematics and supernova remnant associations are also available \revision{(and can provide crucial information for some sightlines; e.g., \citealt{jing2023})}, they tend to have significantly larger uncertainties \citep[e.g.][]{frail1990,verbiest2012}. Previous models necessarily included these more imprecise distance measures due to the small number of pulsar distance measurements available overall; as we will show, the sample of high-precision distances based on parallax and globular cluster associations is now sufficiently large to rely exclusively on these measures.

In Appendix Table~\ref{tab:parallaxes}, we list 126 pulsar parallaxes used in this work, including parallaxes based on VLBI, timing, and optical counterparts in Gaia \citep[e.g.][for complete references, see Table~\ref{tab:parallaxes}]{2016AA...587A.109G,2018ApJ...864...26J,deller2019,moran23}.\footnote{Many of these parallaxes can be found in a database maintained by Shami Chatterjee at \url{https://hosting.astro.cornell.edu/research/parallax/}.} To reduce model-fitting errors, all parallaxes are required to have $<25\%$ fractional uncertainty for inclusion in our analysis. The selection of timing parallaxes requires additional care. Timing parallaxes suffer from different systematic uncertainties than VLBI,  including pulsar spin noise in timing measurements and a dependence on ecliptic latitude \citep[e.g.][]{mpta-parallaxes2024,fiscella2025}. In some cases, reported timing parallaxes for an individual pulsar can change significantly between datasets of different time spans, even when the measurements are reported with high significance \citep{nanograv-9yr,nanograv-12yr}. In cases where a timing parallax has been measured multiple times, whether in independent studies or in different iterations of a multi-year timing dataset, we therefore require that the parallax has remained consistent to within $1\sigma$ across those datasets for inclusion in our sample. 

At the time of our analysis, 45 globular clusters are known to contain pulsars \revision{(for a review of recent discoveries and search strategies, see, e.g., \citealt{das2025a,das2025b,padmanabh2024,ridolfi2022,pan2021_globs})}.\footnote{A database is maintained by Paulo Freire at \url{https://www3.mpifr-bonn.mpg.de/staff/pfreire/GCpsr.html}.} These globular clusters are listed in Appendix Table~\ref{tab:globs}, with distances from \cite{baumgardt2021}, who derive them primarily from Gaia parallaxes or kinematically through a combination of Gaia and Hubble Space Telescope velocity data. All globular cluster distances used in this work have $<7\%$ fractional uncertainty, and the majority have $<2\%$ fractional uncertainty. Many of these clusters contain multiple pulsars with small variations in DM due to the pulsars' different spatial positions within a given cluster. In these cases, we adopt the mean DM of all pulsars within a cluster. In total, our sample consists of 171 distances.

Figure~\ref{fig:2001-2025-comparison} shows the all-sky distribution of distances used in this work, compared to all distance inputs used to calibrate NE2001. There are $\approx4\times$ more distances at high Galactic latitudes ($|b|>20^{\circ}$), providing significantly improved coverage of the thick disk. The number of parallax measurements alone now outnumbers all distances used to calibrate NE2001, and when combined with globular clusters distances, is comparable to the total number of distances used to calibrate YMW16, many of which were less precise. A key limitation of the present-day parallax distance sample is the dearth of measurements in the Southern hemisphere (visible as a gap at Galactic longitudes $\lesssim -60^\circ$ in Figure~\ref{fig:2001-2025-comparison}). The overall asymmetry in the spatial distribution of precise pulsar distances used in this work should be taken into account when considering electron density model uncertainties. 

\begin{figure}
    \centering
    \includegraphics[width=0.43\textwidth]{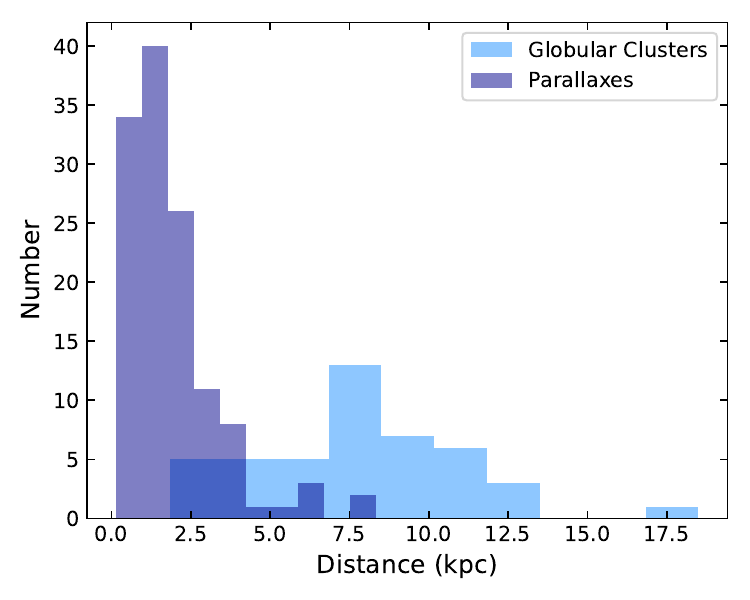}
    \caption{Histogram of distance measurements used in this work. Globular clusters (light blue) provide the largest distances in the sample, whereas parallaxes (dark blue) predominantly yield distances $<3$ kpc.}
    \label{fig:distance-hist}
\end{figure}

Figure~\ref{fig:distance-hist} compares the distribution of distances probed by parallaxes and globular clusters. While pulsar parallaxes are largely biased to probing $<3$ kpc from the Sun, globular clusters probe a broad range of distances from $\approx 2$ kpc to $18$ kpc, with the most distant globular clusters in the sample residing at high Galactic latitudes. 
In many cases, globular clusters are several thick-disk scale heights above the mid-plane, providing strong constraints on the asymptotic DM for individual LOSs. Model uncertainties are distance-dependent, and exacerbated for sightlines more than a few kiloparsecs from the Sun (particularly in the Galactic plane).

\subsection{Comparison to Model-Predicted Distances}

\begin{figure*}
    \centering
    \includegraphics[width=0.8\textwidth]{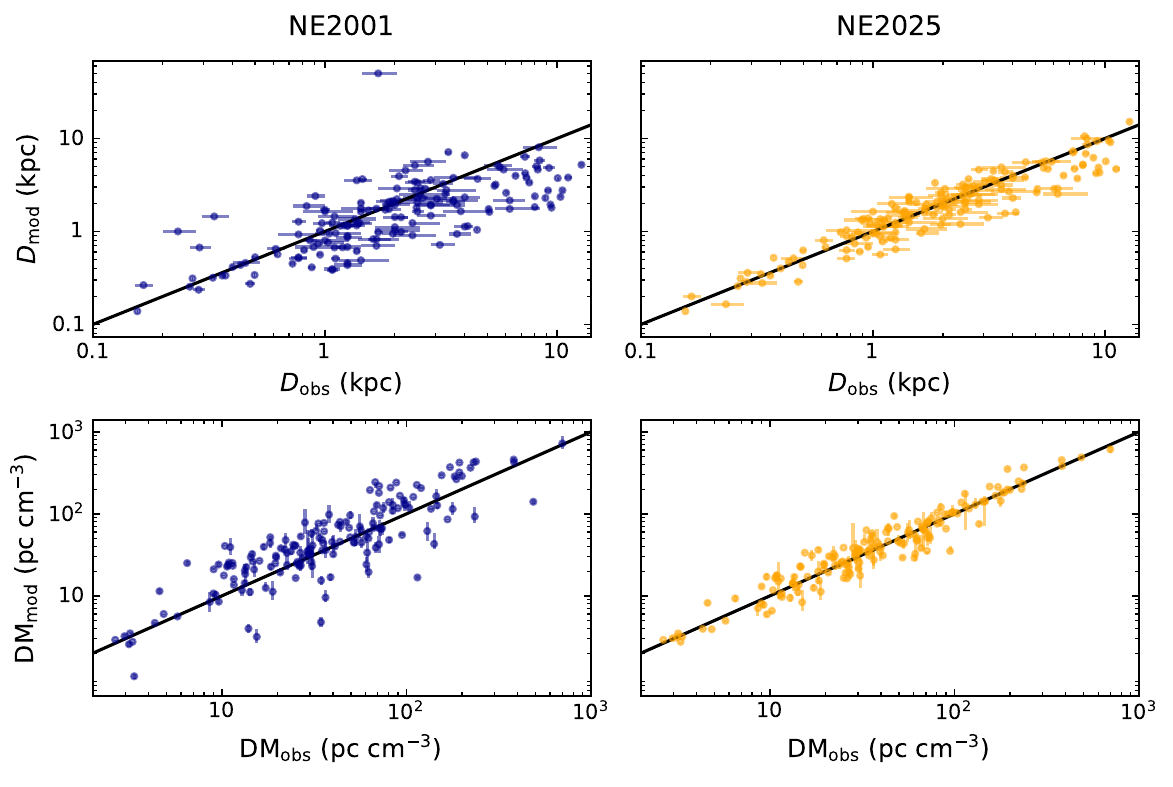}
    \caption{Comparison between predicted and observed distances (top row) and DMs (bottom row) for NE2001 (left; purple points) and NE2025 (right; orange points). Black diagonal lines indicate where the predictions match the measurements.}
    \label{fig:dist-diff}
\end{figure*}

We use the independent distance sample to identify components of NE2001 that require significant revision, by comparing measured and predicted distances and DMs. LOSs more than 4 kpc above the Galactic plane are excluded when comparing predicted and observed distances, 
because small differences between model and actual DMs can produce large distance errors as the DM perpendicular to the disk (i.e. ${\rm DM_{\perp} = DM} \sin \vert b \vert$) approaches an asymptotic value as a function of distance. Figure~\ref{fig:dist-diff} shows a comparison between the measured distances and the distances predicted by NE2001, as well as a comparison between measured and predicted DMs for the entire distance sample. The same comparison is made for NE2025. 
The majority of distances are underestimated by NE2001. This trend manifests in the DM distribution as a systematic overestimation of pulsar DM based on distance -- i.e., the model has an overabundance of free electrons for these LOSs, and hence reaches the observed DM at a smaller distance than is observed. This systematic overestimation of DM (and underestimation of distance) affects LOSs across a broad range of observed distances and spatial coordinates, suggesting that a reduction in density for a single disk component is required. This problem is attributable to the thick disk, and is ameliorated by refitting that model component (see Section~\ref{sec:thick}). 

For the minority of cases where the distance is overestimated by NE2001, the discrepancy between the predicted and observed distances is significantly larger than for the underestimated distances. There are twelve pulsars with predicted distances that are $>100\%$ larger than the measurements. In these cases, predicting the DM based on distance yields a DM significantly lower than observed -- i.e., the model has a deficiency of free electrons along the LOS. We compared these pulsars to the locations of known \HII regions, using methods and selection criteria discussed in \cite{ocker2024c}. Based on those criteria, five of these pulsars appear to be spatially coincident with \HII regions. We therefore attribute these model discrepancies to discrete ISM overdensities, which are included as new clumps in the updated model (Section~\ref{sec:clumps}).  

Arguably some of the most complex ISM structure explicitly modeled in NE2001 is the local ISM, which includes the Local Bubble, Loop I, a local superbubble and low-density region, as well as the Gum Nebula and the Vela supernova remnant. The model performance of these local ISM components can be seen in Figure~\ref{fig:dist-diff} by comparing the predicted and observed distances of pulsars within 700 pc of the Sun. Overall, the difference between observed and predicted distances is small ($<50\%$) for most of these nearby pulsars. Three pulsars show a $>100\%$ discrepancy due to unmodeled overdensities along the LOS; these are included as new clumps. The Vela pulsar, J0835$-$4510, additionally has a large discrepancy in DM, ${\rm DM_{NE2001} - DM_{obs}} = 176$ pc cm$^{-3}$. Notably, the discrepancy disappears if the Vela supernova remnant is placed closer to the observed distance of the pulsar, thus motivating a revision to the distance to Vela (discussed further in Section~\ref{sec:local}).

\subsection{Comparison to Dispersion Measures of the Pulsar Population}

\begin{figure*}
    \centering
    \includegraphics[width=0.85\textwidth]{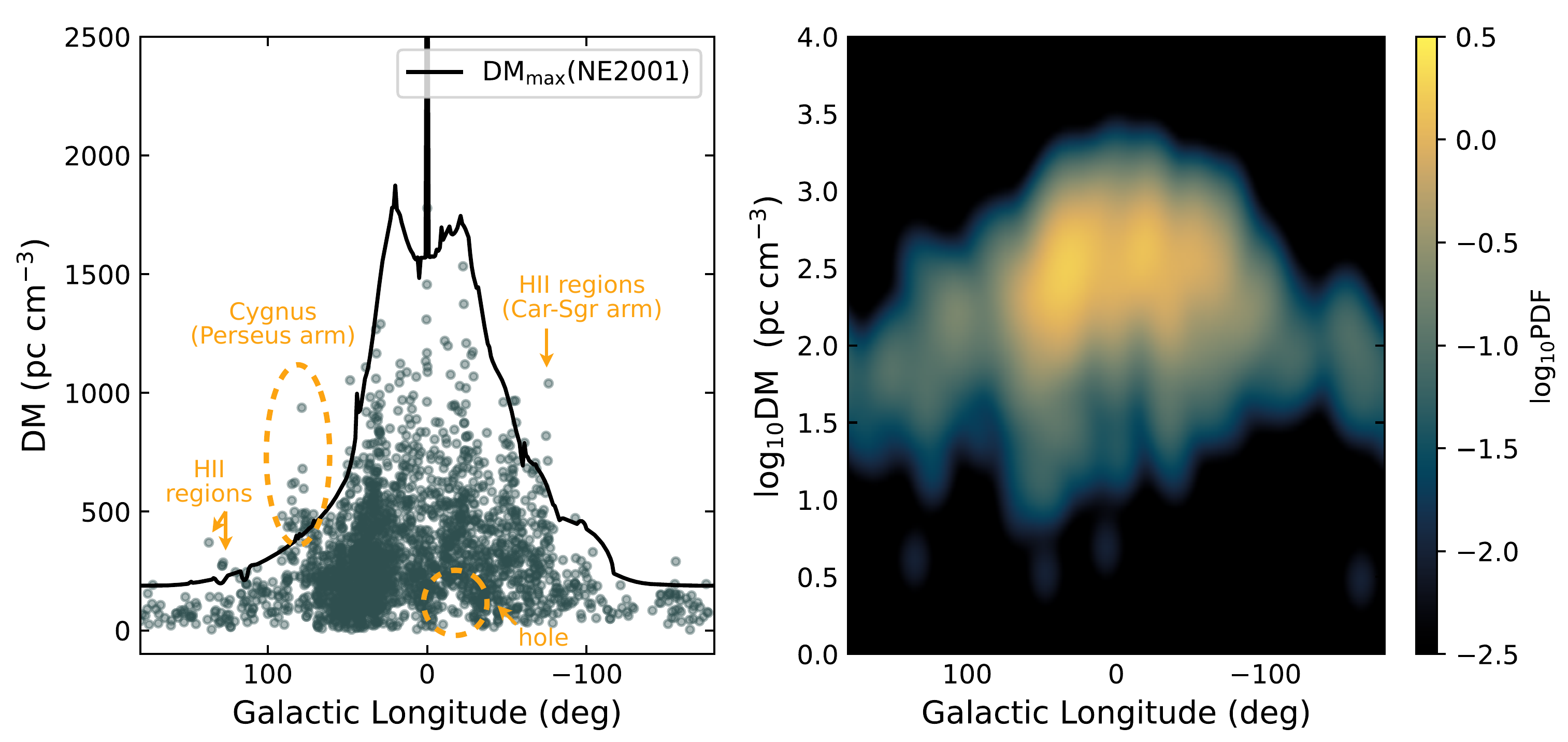}
    \caption{Left: Dispersion measure vs. Galactic longitude for all Galactic pulsars at latitudes $|b|<5^\circ$. The maximum DM predicted by NE2001 is shown by the black curve, and is evaluated at a resolution of $1^\circ$ in longitude for $b=0^\circ$. Specific regions of interest for the updated model, including \HII regions and a hole at $l\approx-20^\circ$, are highlighted in orange. Right: A Gaussian kernel-density estimator (KDE) of the probability distribution function for the DMs shown in the righthand panel. The KDE is evaluated using logarithmic bins in DM, \revision{and highlights the significance of the hole noted in the lefthand panel, which is revisited in Section~\ref{sec:clumps}.}}
    \label{fig:dm-vs-lon1}
\end{figure*}

In addition to pulsars with known distances, we examine how NE2001 compares to the DM distribution of the entire known pulsar population, based on v2.7 of the ATNF Pulsar Catalog \citep{psrcat}.\footnote{\url{https://www.atnf.csiro.au/research/pulsar/psrcat}} Figure~\ref{fig:dm-vs-lon1} shows DM vs. Galactic longitude for all known, low-latitude pulsars, as well as a kernel-density estimator of the probability distribution function (PDF) for the same data. Two features are readily apparent in this figure: First, several pulsars have DMs larger than the maximum prediction of NE2001, and the model spuriously predicts them to be at least 50 kpc away. In \cite{ocker2024c}, we showed that nearly all of these pulsars with DMs in excess of the NE2001 maximum are behind known \HII regions. Many of these pulsars are located in the direction of the Cygnus region, which is already included in NE2001 as a set of clumps that produce significant scattering but little DM. The second feature highlighted in Figure~\ref{fig:dm-vs-lon1} is a hole in low DMs at $l\approx-20^\circ$, which is also visible as a decrease in the DM PDF at the same longitudes. This hole is apparent in studies spanning different pulsar surveys and has persisted over decades (see e.g., Figure 11 in \citealt{ne20011}).  This feature could signify an actual  dearth of pulsars in these low-latitude directions but it is more likely  related to the underlying electron density in a volume relatively close to the Sun.\footnote{It is difficult to identify any survey selection bias that would yield such a gap in the DM distribution, given that low-DM pulsars tend to be closer to the observer and therefore easier to detect.} A natural explanation for the presence of such a gap in the DM distribution is an ISM {\it overdensity}, the potential nature of which we discuss further in Section~\ref{sec:clumps}.

\subsection{Comparison to Observed Scattering}

NE2001 predicts scattering by forward modeling the scattering measure ${\rm SM} = \int_0^D C_{\rm n}^2 dl$, where $C_{\rm n}^2$ is the amplitude of the density fluctuation power spectrum, which is then related to scattering observables including the pulse broadening delay $\tau$, angular diameter $\theta_{\rm d}$, and scintillation bandwidth $\Delta \nu_{\rm d}$ by assuming a Kolmogorov spectrum of density fluctuations. The SM and DM are related by
\be
d\DM &=& \nelec ds
\\
d\SM &=& C_{\rm SM} \Fc \nelec^2 ds,
\ee
where the density fluctuation parameter $F_c$ designates the scattering strength of a given model component, and is larger for media that produce more scattering per unit $\rm DM^2$. The total predicted scattering thus depends on an integral over the electron density squared, $F_c$, and the location of each model component along the LOS. Clumps with nonzero $F_c$ along the LOS often tend to dominate the total predicted scattering. 

\begin{figure*}[ht!]
    \centering
    \includegraphics[width=0.8\textwidth]{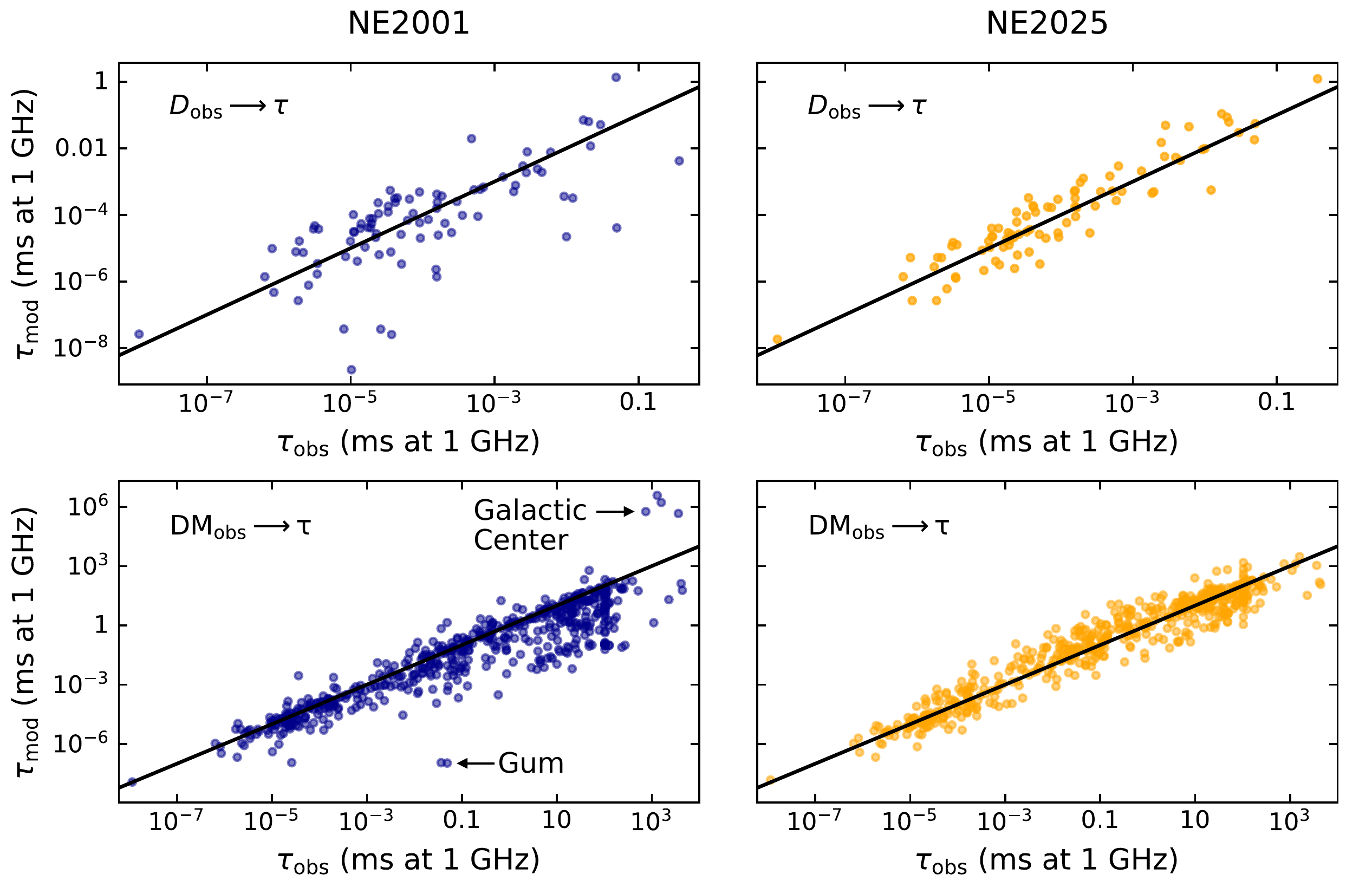}
    \caption{Comparison between predicted and observed scattering delays ($\tau$) for NE2001 (left) and NE2025 (right). The top row shows the result of running the models with the observed distance as input, and the bottom row shows the result of running the models with the observed DM as input. Diagonal black lines indicate where predictions match measurements. Only a subset of pulsars with scattering measurements also have known distances, resulting in the different sample sizes between the top and bottom panels. Two sets of extreme outliers are labeled in the bottom left panel, corresponding to LOSs towards the Galactic Center and the Gum Nebula.}
    \label{fig:tau-diff}
\end{figure*}

We test NE2001's scattering predictions by comparing them to 568 published pulsar scattering measurements, including those compiled by \cite{cordes2016,cordes2022} and about 100 pulse broadening measurements from the FAST Galactic Plane Pulsar Survey \citep{jing2025}. 
We compare the observations to the model in two ways: For the subset of $\approx 90$ pulsars with precise distance measurements and published scattering, we evaluate NE2001 using the distance as input. We also evaluate NE2001 for the entire sample of 568 pulsars, using the observed DM as input. 

The results of this comparison are summarized in Figure~\ref{fig:tau-diff}. When we use the measured distance as input to NE2001, \revision{we find that $53\%$ of scattering times are overestimated}, and typically by less than an order of magnitude, which is comparable to the empirical spread in $\tau$ that is observed for pulsars at a given DM \citep[e.g.][]{rickett70,cordes91,lewandowski2015,krishnakumar15,cordes2022}.
We identify nine outliers where $\tau$ is significantly underestimated by the model for their observed distances, and six of these have distances that are dramatically overestimated by the model based on their DMs. We conclude that a deficiency of free electrons along the modeled LOS is responsible for these cases where the scattering is underestimated from the observed distance, and we use the observed $\tau$ to constrain new clumps described in Section~\ref{sec:clumps}.

When we evaluate NE2001's scattering predictions using the observed DM as input, we find that $\tau$ tends to be underestimated for the majority of sightlines. This underestimation becomes more severe for large observed scattering delays, which correspond to LOSs that predominantly trace the inner Galaxy. As we will show in Section~\ref{sec:spiral}, the underestimation of scattering along these sightlines is corrected by revisions to the thin disk and spiral arms. We also identify two sets of extreme outliers: Four pulsars have $\tau$ drastically overestimated from DM, all of which are LOSs towards the Galactic Center ($|l| < 0.2^\circ$, $|b|<0.3^\circ$). Three pulsars have $\tau$ drastically underestimated from DM and are in the direction of the Gum Nebula. These results motivate revisions to the Gum Nebula and Galactic Center described in Sections~\ref{sec:local} and \ref{sec:GC}, respectively.

\section{Fitting Criteria}\label{sec:fitting}

\revision{In the previous section, we found that specific model components are primarily responsible for discrepancies between NE2001 predictions and observations. We therefore proceed by refitting these individual model components, using the various datasets described in Section~\ref{sec:model-comparison}.} Model components are fit based on three main criteria:

\begin{enumerate}
    \item Minimization of the difference between observed and predicted distances and DMs. Where appropriate, we use a likelihood analysis and/or $\chi^2$ minimization to fit the parameters of specific model components (e.g., see Section~\ref{sec:thick} and Appendix~\ref{app:fitting-functions}). Due to the limited spatial distribution of the pulsar distance sample (Figures~\ref{fig:2001-2025-comparison}-\ref{fig:distance-hist}), not all model components are constrained by this criterion.
    \item Minimization of the difference between observed and predicted scattering measures. Both observed distances and DMs are used to predict scattering delays $\tau$. We evaluate the absolute value of the mean, median, and rms logarithmic difference between observed and predicted scattering ($\rm log_{\rm 10}(\tau_{\rm mod}/\tau_{\rm obs})$), and seek to minimize all three of these statistical metrics (e.g., see Section~\ref{sec:spiral} and Appendix~\ref{app:fitting-functions}).
    \item Minimization of the number of pulsars with DMs larger than asymptotic
    model values, $\Nxgal$ (excluding the Magellanic Clouds). Some pulsars have observed DMs greater than the model's maximum value integrating through the entire set of model components. While a significant fraction of canonical pulsars can escape the Galaxy \citep{1994Natur.369..127L,cordes_chernoff98,arzoumanian2002} and travel significant distances (a few kpc) during their $\sim 10$\,Myr lifetimes, the expected number of observed pulsars in the Galactic halo is close to zero for existing pulsar surveys \citep{rajwade2018}. We evaluate the second derivative of $N_{\rm >mod}$ as a function of a given model parameter $x$, and adopt the value of $x$ for which $d^2N_{\rm >mod} /dx^2$ is maximized, i.e., the point at which increasing $x$ yields diminishing returns in reducing $N_{\rm >mod}$. This criterion is primarily constraining for the thin disk and spiral arms; see Section~\ref{sec:spiral} for details.
\end{enumerate}

The criteria described above are similar to those used to optimize NE2001; as discussed in \cite{ne20012}, criterion (3) is equivalent to reducing the number of pulsars with anomalous predicted luminosities. The specific fitting functions used in the criteria above are elaborated upon in Appendix~\ref{app:fitting-functions}. \revision{An alternative approach to the fitting methodology used here would be to perform a global fit to all model components simultaneously. Our approach to fitting model components individually likely introduces additional systematic uncertainties, and alternative fitting strategies will be explored further in future work.}

Not all of our fitting criteria yield well-defined uncertainties for individual parameters.  Actual errors exceed formal fitting errors  because for some the observables have non-Gaussian distributions and others are subject to systematic uncertainties.  Only criterion (1) explicitly accounts for measurement uncertainties in observed distances. The very small measurement uncertainties in DM, typically $<10^{-2}$ pc cm$^{-3}$, have negligible effect on our analysis. Scattering time measurement uncertainties are difficult to quantify because they depend on the pulse broadening function (PBF) adopted when fitting for scattering delays $\tau$ \citep[e.g.][]{geiger2024}, yielding sizable systematic errors in estimates of $\tau$.  Nonetheless, in most cases where we optimize model parameters based on comparing predicted and observed scattering distributions, the spread in measurement values is much larger than the uncertainty on any given measurement. For these reasons, few of the parameter values adopted in \model\ have statistically meaningful formal errors. For the spiral arms, model parameter uncertainties are primarily due to finite quantization of the parameter grid space used. Where relevant, we discuss the interpretation of specific model parameter uncertainties in subsequent sections. 

Given the considerations above and the covariances between a number of model parameters, it is difficult to define a robust uncertainty on the model prediction for any given LOS. The most straightforward metric for quantifying uncertainties in the model predictions is by comparing observation to prediction, and we discuss the \model\ model performance in this context in Section~\ref{sec:conc}. However, it is important to note that, as with all Galactic electron density models, \model\ does not represent a unique solution for the ISM electron density -- given the model is under-constrained by observations, certain model components require a priori assumptions about, e.g., their analytic form, often based on observational constraints at other wavelengths (e.g., adopting spiral arm locations traced by \HII regions). For the purposes of this model update, all a priori assumptions about Galactic structure are identical to those used in NE2001.

\section{Thick Disk}\label{sec:thick}

The thick disk component of NE2001 is given by
\begin{gather}
    n_1 G_1(r,z) = n_1g_1(r)h_1(z/H_1) \\
    g_1(r) = [{\rm cos}(\pi r/2A_1)/{\rm cos}(\pi R_\odot/2A_1)]U(A_1-r) \\
    h_1(z/H_1) = {\rm sech^2}(z/H_1)
\end{gather}
where $n_1$ is the mid-plane density, $g_1(r)$ describes the radial shape of the disk, which is cut off by the unit step-function $U(r)$ at a radial distance $A_1 = 17.5$ kpc, and $h_1(z/H_1)$ describes the shape of the disk transverse to the Galactic plane, with a scale height $H_1$ \citep{ne20011}. As with all model components, we use a Galactocentric reference frame. We refit only the vertical ($z$-dependent) parameters of the thick disk, $n_1H_1$ and $H_1$, by focusing on \revision{57} sightlines dominated by this component -- namely, pulsars that are at high Galactic latitudes ($|b|>20^\circ$) and that are not extremely discrepant from NE2001 (due, e.g., to unmodeled clumps along the LOS). Given the restriction to high-latitude sources, the local ISM contributions in the model are similar between sightlines. 

We perform a grid search for $(n_1H_1, H_1)$ using two methods: (1) A log-normal probability density function is used to construct the likelihood function for $(n_1H_1, H_1)$ given observations of distance (and its uncertainties) and DM (based on methods in \citealt{ocker2020}; see Appendix~\ref{app:fitting-functions}); and (2) simple $\chi^2$ minimization of the linear difference between observed and predicted DM. Both methods fit for the predicted DM based on observed distance, in order to include high-$|z|$ globular clusters that constrain the asymptotic value of ${\rm DM_\perp} = {\rm DM sin}|b|$, which is $\lesssim n_1H_1$ for LOSs that sample only the thick disk. \revision{We fit for the product $n_1H_1$ because it is directly constrained by these high-$|z|$ sightlines, and because $n_1$ and $H_1$ are highly covariant as individual parameters.}

\begin{figure*}
    \centering
    \includegraphics[width=0.8\textwidth]{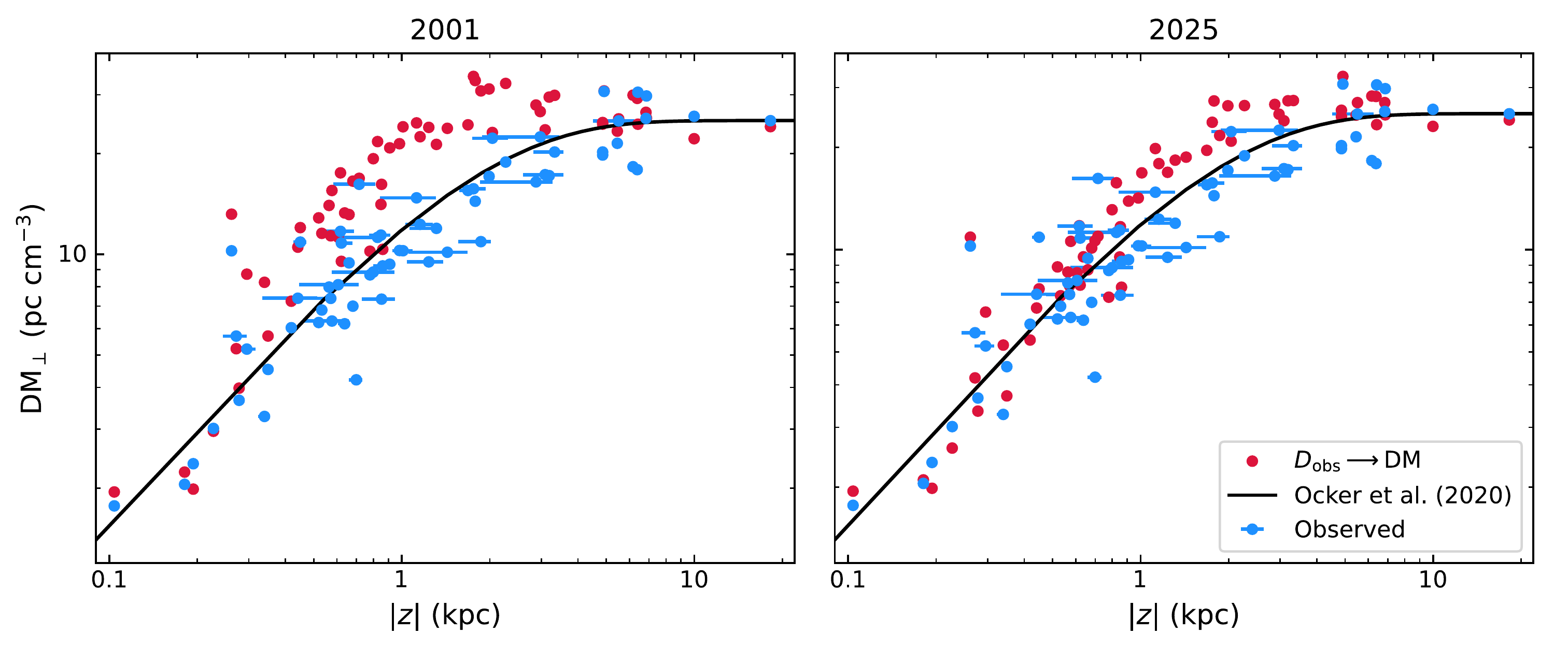}
    \caption{Dispersion measure perpendicular to the Galactic plane (${\rm DM_
    \perp = DM\ sin}|b|)$ vs. $|z| = D\ {\rm sin}|b|$, the distance perpendicular to the Galactic plane  for high-latitude pulsars ($|b|>20^\circ$) with measured distances in blue and model-based distances in red. The left hand panel shows NE2001 predictions, while the right hand panel shows predictions based on the refitted thick disk. We have excluded pulsars identified as outliers in Section~\ref{sec:model-comparison}, in order to isolate the thick disk component of the model. The best-fit exponential thick disk model from \cite{ocker2020} is shown in black.}
    \label{fig:dmperp-vs-z}
\end{figure*}

Both methods yield consistent best-fit values of ($n_1H_1,H_1$) with similar uncertainties. However, adopting the best fit from $\chi^2$-minimization, $n_1H_1 = 0.024\pm0.004$ kpc cm$^{-3}$ and $H_1 = 1.59\pm0.60$ kpc, leaves $\sim 100$ pulsars \revision{in the broader known population (without independent distances)} at high latitudes that have observed DMs greater than the maximum prediction of the model. These pulsars are not segregated in longitude or latitude, ruling out a clear relationship to alternative structure for the thick disk (including a warp). Increasing $n_1H_1$ to its $+1\sigma$ value, $0.028$ kpc cm$^{-3}$, removes nearly all of these discrepant pulsars, and so we set $n_1H_1 = 0.028$ kpc cm$^{-3}$ in the final version of the model. An alternative approach could invoke a cost function in the likelihood analysis that explicitly penalizes parameter values that yield a DM deficit; regardless, we expect the best-fit results to be similar, \revision{and alternative fitting strategies will be pursued in future work}.
The errors quoted here on $n_1H_1$ and $H_1$ are the formal $1\sigma$ fitting uncertainties, which do not account for the uncertainties implicit to all of the other NE2001 parameters held fixed in the grid search. With the revised ($n_1H_1, H_1$), we find $F_c = 0.18$, the fluctuation parameter used for the thick disk in NE2001, provides an adequate fit to the observed scattering delays for the same subset of high-latitude pulsars, and $F_c$ is left unchanged in the updated model.

Figure~\ref{fig:dmperp-vs-z} shows $\rm DM_\perp$ vs. $|z| = D\ {\rm sin}|b|$ for the sample of high-latitude pulsars used to fit the model, compared to NE2001 and our refitted thick disk. In NE2001, $n_1H_1 = 0.033$ kpc cm$^{-3}$ and $H_1 = 0.97$ kpc.\footnote{Here and elsewhere we quote NE2001 parameter values as implemented in the final released version, which differ slightly from the values quoted in \cite{ne20011,ne20012}.} Reducing $n_1H_1$ and increasing $H_1$ mitigates NE2001's tendency to overestimate $\rm DM_\perp$ at a given $|z|$, and our new best-fit parameters significantly close the gap between observed and predicted DM for these high-latitude pulsars. Figure~\ref{fig:dmperp-vs-z} also shows the best-fit exponential model for the thick disk from \cite{ocker2020}, who found $\rm DM_\perp(\infty) = 23.5 \pm 2.5$ pc cm$^{-3}$ and $H_1 = 1.57\pm0.15$ kpc, comparable to the values we found using a ${\rm sech^2}(z)$ model \revision{(and similar to values found by \citealt{berkhiujsen,2008PASA...25..184G,2009ApJ...702.1472S,2012MNRAS.427..664S})}. Notably, the ${\rm sech^2}(z)$ distribution leads to a sharper rise in $\rm DM_\perp$ vs. $|z|$, which leads to a slight overestimation of $\rm DM_\perp$ at $|z|\approx 1.5$ kpc (around the scale height). The exponential distribution adopted by \cite{ocker2020} and many previous studies (e.g., \citealt{2008PASA...25..184G,2012MNRAS.427..664S}; and refs. therein) has a shallower rise in $\rm DM_\perp$ vs. $|z|$, and appears to be more consistent with observations (albeit with significant scatter). While NE2001 adopts a ${\rm sech^2}(z)$ distribution in order to avoid a non-physical, exponential rise in electron density for LOSs approaching $b = 0^\circ$, an exponential distribution for the thick disk is worth exploring in future models.

\section{Spiral Arms \& Thin Disk}\label{sec:spiral}

\begin{figure}
    \centering
    \includegraphics[width=0.45\textwidth]{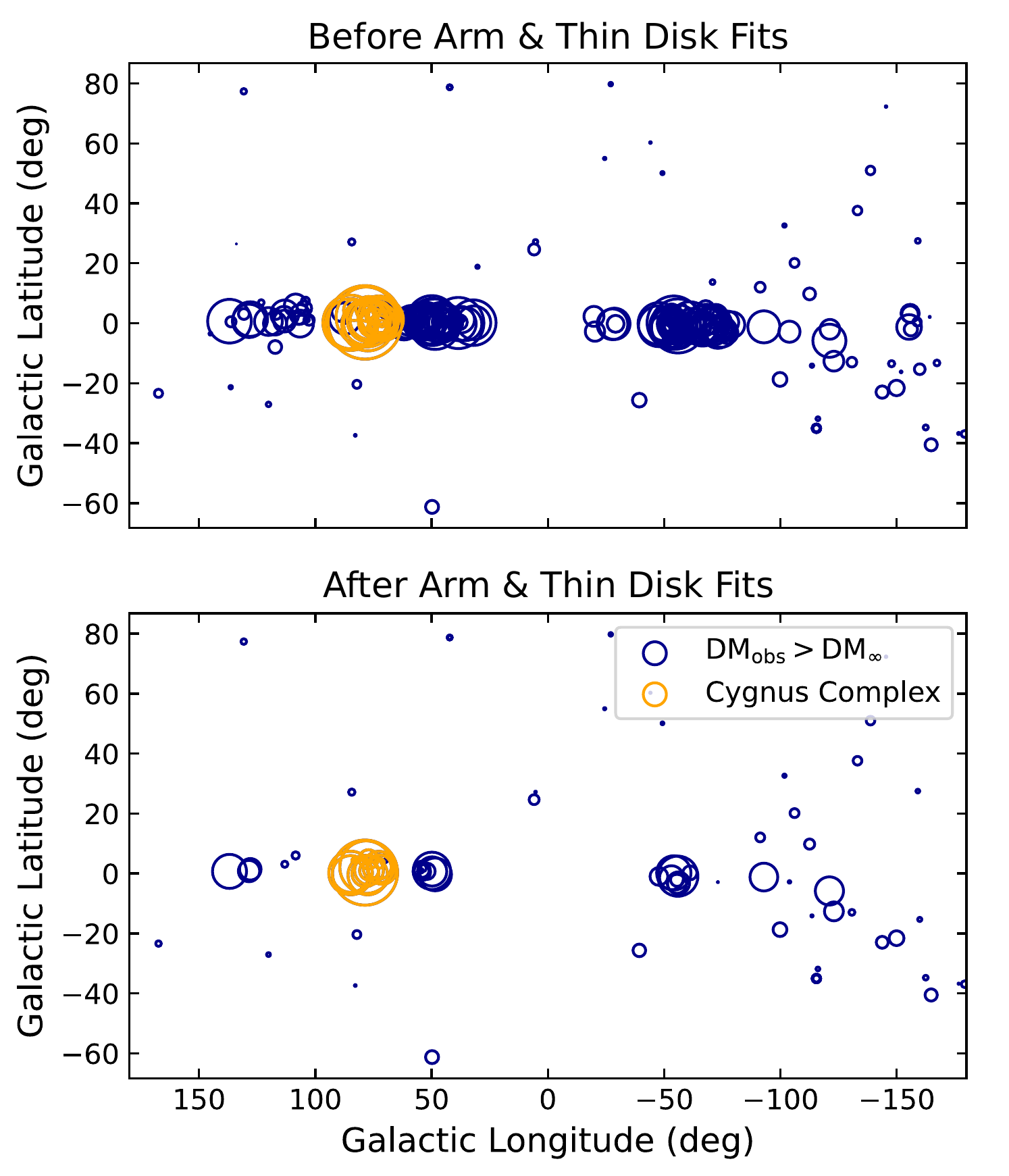}
    \caption{The sky distribution of pulsars with observed DMs ($\rm DM_{obs}$) greater than the maximum model values along their LOSs ($\rm DM_\infty$), before (top) and after (bottom) changing the spiral arms and thin disk of NE2001. The top panel shows pulsars with $\rm DM_{obs} > DM_{\infty}$, after re-fitting only the thick disk component of NE2001 (Section~\ref{sec:thick}). The bottom panel shows pulsars with $\rm DM_{obs} > DM_{\infty}$ after re-fitting the thin disk and arms 2, 3, and 4 (Section~\ref{sec:spiral}). The sizes of the points correspond to the amount of observed DM in excess of the model maximum; larger points indicate larger excess DM. The Cygnus complex is highlighted in orange and is treated separately from the spiral arm and thin disk fitting.}
    \label{fig:before-after-arms-nxgal}
\end{figure}

The adopted thick disk yields a significant reduction in the mid-plane density: $n_1$ decreases by nearly a factor of two from $0.035$ cm$^{-3}$ to $0.018$ cm$^{-3}$. This reduction leaves $N_{\rm >mod}\approx 180$ pulsars \revision{in the entire known population} with DMs greater than the model's maximum prediction \revision{(none of which have known distances)}. The majority of these pulsars are concentrated in the Galactic plane with $\vert b \vert < 10^{\circ}$(see Figure~\ref{fig:before-after-arms-nxgal})\footnote{These numbers exclude pulsar sightlines intersecting the Cygnus and W51 complexes, which we treat separately to avoid biasing the spiral arm parameters.}, indicating that the spiral arm and thin disk components require higher electron density columns to account for the lower density of the thick disk.

\revision{NE2001 uses logarithmic spiral arms adapted from the parameterization in \cite{wainscoat92}, and given by:}
\begin{gather}
    n_a G_a(\mathbf{x}) = n_a \sum_j f_j g_{a_j}(r,\mathbf{x})h(z/h_jh_a)\\
    g_{a_j}(r,\mathbf{x}) = e^{-s_j(\mathbf{x}/w_jw_a)^2} \{{\rm sech^2}\bigg(\frac{r-A_a}{2}\bigg)U(r-A_a)\}\\
    h(z/h_jh_a) = {\rm sech^2}(z/h_jh_z)
\end{gather}
\revision{where $(n_a,w_a,h_a)$ are constants describing arm peak density, width, and height, and are multiplied by weight factors $(f_j,w_j,h_j)$ for each arm. The Galactocentric location of each arm in the plane is given by $s_j(\mathbf{x})$, and the arms are truncated by the unit step function $U(r)$ at a constant $A_a$ in Galactocentric radius $r$.}

\revision{The thin disk is given by}
\begin{gather}
    n_2G_2(r,z) = n_2g_2(r) h(z/H_2)\\
    g_2(r) = e^{-[(r-A_2)/1.8]^2}U(10\ {\rm kpc} - r)\\
    h_2(z) = {\rm sech^2}(z/H_2)
\end{gather}
\revision{where $n_2$ is the peak mid-plane density, $A_2$ is the radius at which the thin disk density peaks, and $H_2$ is the scale height of the thin disk. The thin disk is truncated by the unit step function $U(r)$ at a radius of 10 kpc, although the thin disk density drops to $<10\%$ of its peak value beyond a radius of 7 kpc.}

We evaluate the gradient of $N_{\rm >mod}$ \revision{for the aforementioned 180 pulsars} vs. \revision{the weights given to each arm's peak density, width, and height $(f_j,w_j,h_j)$, as well as the peak mid-plane density, scale radius, and scale height of the thin disk ($n_2,A_2,H_2)$.} We identify five parameters that have a significant effect on $N_{\rm >mod}$: the thin disk radius ($A_2$), the \revision{density weights of arms 2, 3, and 4 ($f_{2}, f_{3}, f_{4}$), and the weighting of the arm 4 width ($w_4$).}
Based on fitting criterion (3) -- see Section~\ref{sec:fitting} -- we identify the optimal values of these five parameters, which are listed in Table~\ref{tab:large-component-values}. \revision{The constant multipliers of each arm's density, width, and height ($n_a, w_a, h_a$) are left fixed at their original values.} Relative to NE2001, the thin disk radius increases from 3.8 kpc to 4.3 kpc, and is still in line with observations of the molecular ring \citep{urquhart2014,anderson2015}. The central densities of arms 3 and 4 (the Car-Sgr and Perseus arms) increase substantially, by a factor of two for arm 3 and a factor of 3.7 for arm 4. These revisions leave $N_{\rm >mod} \approx 40$ pulsars, the majority with DM excess $<9$ pc cm$^{-3}$; see Figure~\ref{fig:before-after-arms-nxgal}. There are several pulsars remaining in the plane with substantial DM excess ($>100$ pc cm$^{-3}$) that are likely related to discrete structures along the LOS; these pulsars are treated separately in Section~\ref{sec:clumps}. 

\begin{figure}
    \centering
    \includegraphics[width=0.45\textwidth]{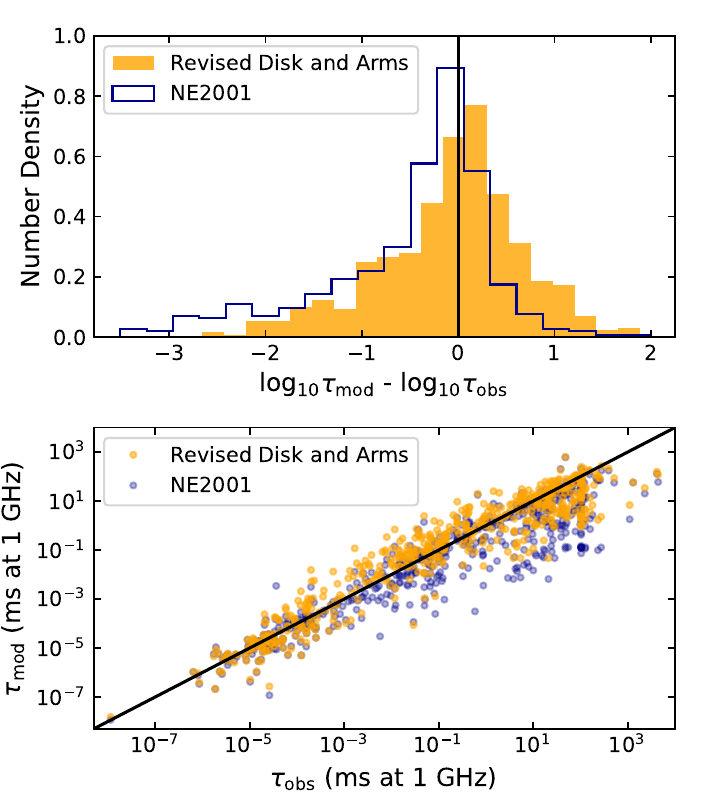}
    \caption{Pulse broadening predicted from DM, before and after changing the spiral arms and thin disk of NE2001. The top panel shows normalized histograms of the logarithmic difference between predicted and observed scattering delays $\tau$, in dark blue for NE2001 and in orange for a model with the spiral arm and thin disk parameters revised based solely on minimizing $\Nxgal$ (\revision{prior to refitting the fluctuation parameters;} see Section~\ref{sec:spiral}). The bottom panel shows the predicted vs. observed scattering delays for the same two model setups. Black lines indicate $\tau_{\rm mod} = \tau_{\rm obs}$ in both panels. For the purposes of this comparison we exclude extreme outliers in scattering (the Galactic Center and Gum Nebula), which are treated separately.}
    \label{fig:before-after-arms-taudiff}
\end{figure}

These revisions to the spiral arms and thin disk have substantial impact on the scattering predictions of the model. In Figure~\ref{fig:before-after-arms-taudiff}, we show the distribution of $\tau_{\rm mod}$ vs. $\tau_{\rm obs}$ for NE2001, and after revising the spiral arms and thin disk based solely on the $N_{\rm >mod}$ criterion \revision{(without any adjustment to the spiral arm and thin disk fluctuation parameters)}. \revision{Even before refitting the fluctuation parameters,} the scattering distribution becomes more nearly symmetric about $(\tau_{\rm mod}/\tau_{\rm obs}) = 1$. Given that these pulse broadening delays are for a sample of pulsars that is largely independent of those used in the $N_{\rm >mod}$ minimization described above, this comparison affirms that our model is self-consistent, despite the lack of model uniqueness discussed in Section~\ref{sec:fitting}. 

We further optimize the fluctuation parameters of the thin disk and spiral arms ($F_2, F_a$) based on the criterion that the absolute value of the median, mean, and rms difference $\rm log_{10}(\tau_{\rm mod}/\tau_{\rm obs})$ should be minimized. This procedure yields a new value for the spiral arm fluctuation parameter $F_a = 3.7$ (slightly lower than in NE2001), and an unchanged value for the thin disk $F_2 = 120$. The resulting $\rm log_{10}(\tau_{\rm mod}/\tau_{\rm obs})$ has a median of $-0.0001$, a mean of $-0.11$, and an rms of $0.76$. This rms log-difference is nearly identical to the empirical spread in the observed $\tau$-DM relation \citep{krishnakumar15,cordes2022}, indicating that \model\ gives scattering predictions that are as robust as the empirical relation.

Figure~\ref{fig:electron-map} shows the model's electron density structure in the Galactic plane, before and after all revisions to the large-scale disk and spiral arm components. The change in density contrast between the thick disk and spiral arms is clearly visible, illustrating the updated partitioning of electrons between the large-scale components.

\begin{deluxetable*}{l C C}\label{tab:large-component-values}
\tablecaption{Parameters of Large-Scale Components for NE2001 and \model}
\tablehead{\colhead{Parameter} & \colhead{NE2001} & \colhead{\model}}
\startdata
\cutinhead{\revision{Thick Disk}}
$n_1H_1$ (kpc cm$^{-3}$) & \mathbf{0.033} & \mathbf{0.028} \\
$H_1$ (kpc) & \mathbf{0.97} & \mathbf{1.59} \\
$A_1$ (kpc) & 17.5 & 17.5 \\
$F_1$ & 0.18 & 0.18 \\
\cutinhead{\revision{Thin Disk}}
$n_2$ (cm$^{-3}$) & 0.09 & 0.09 \\
$H_2$ (kpc) & 0.14 & 0.14 \\
$A_2$ (kpc) & \mathbf{3.8} & \mathbf{4.3} \\
$F_2$ & 120 & 120 \\
\cutinhead{\revision{Spiral Arms}}
$n_af_j$ (cm$^{-3}$) & 0.030\times(0.50,\mathbf{1.2,1.3,1.0},0.25) & 0.030\times(0.50,\mathbf{1.5,2.7,3.7},0.25) \\
$h_ah_j$ (kpc) & 0.25\times(1.0,0.8,1.3,1.5,1.0) & 0.25\times(1.0,0.8,1.3,1.5,1.0) \\
$w_aw_j$ (kpc) & 0.6\times(1.0,1.5,1.0,\mathbf{0.8},1.0) & 0.6\times(1.0,1.5,1.0,\mathbf{0.96},1.0)\\
$A_a$ (kpc) & 11.0 & 11.0 \\
$F_aF_j$ & \mathbf{5}\times(1.1,0.3,0.4,1.5,0.3) & \mathbf{3.7}\times(1.1,0.3,0.4,1.5,0.3) \\
\cutinhead{\revision{Galactic Center}}
$n_{\rm GC0}$ (cm$^{-3}$) & 10.0 & 10.0 \\
$H_{\rm GC}$ (kpc) & \mathbf{0.026} & \mathbf{0.05} \\
$R_{\rm GC}$ (kpc) & 0.145 & 0.145 \\
$F_{\rm GC}$ & \mathbf{6\times10^{4}} & \mathbf{224} \\
\enddata
\tablecomments{Parameter values that are modified in \model\ are highlighted in bold.}
\end{deluxetable*}

\section{Gum Nebula and Vela Supernova Remnant}\label{sec:local}

In Section~\ref{sec:model-comparison}, we identified two key discrepancies in NE2001 predictions of DM and scattering in the direction of the Gum nebula and Vela supernova remnant: (1) The DM of the Vela pulsar is overestimated by $>100\%$, and (2) the scattering of pulsars viewed through the Gum Nebula tends to be underestimated, in some cases by up to $\sim$6 orders of magnitude.

\begin{figure*}
    \centering
    \includegraphics[width=0.9\textwidth]{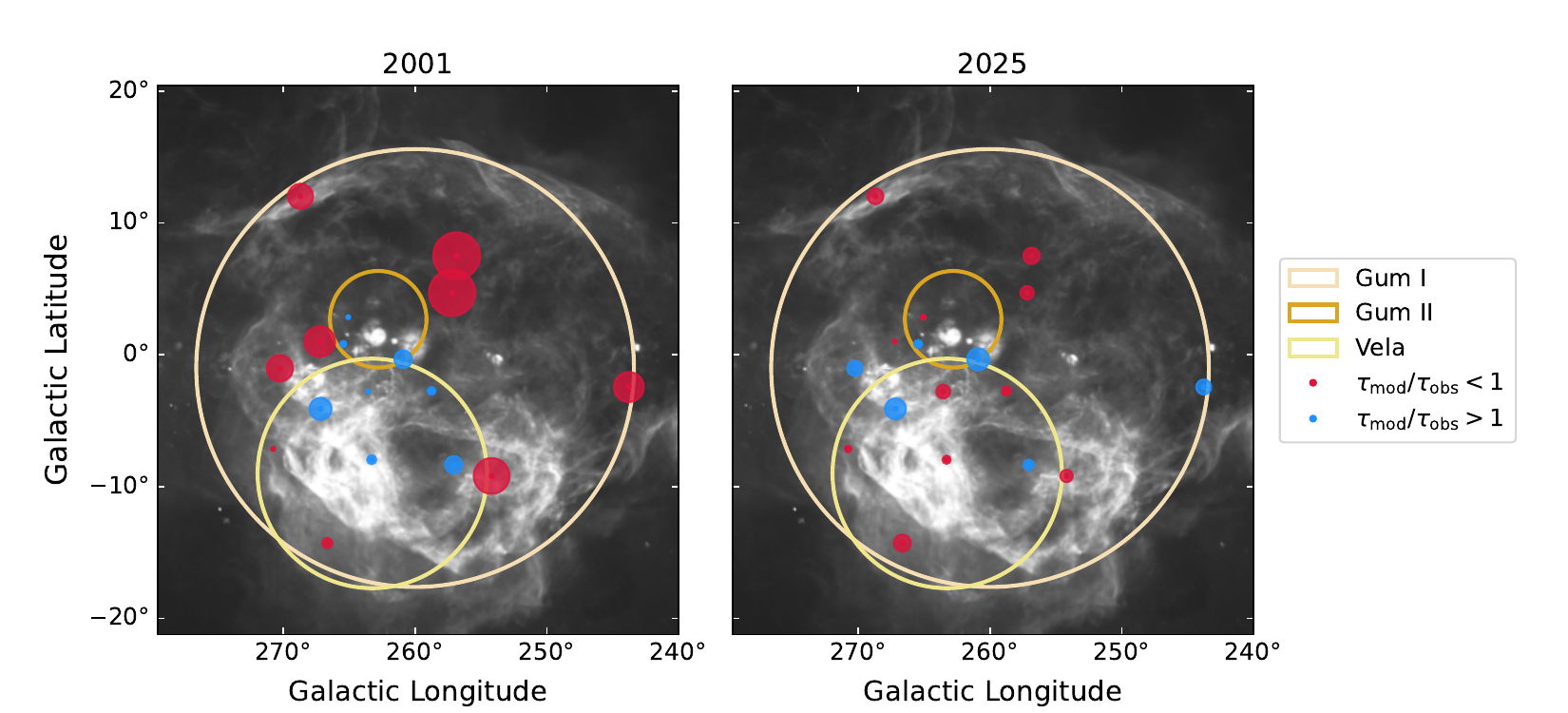}
    \caption{Comparison between observed and predicted scattering of 17 pulsars viewed through the Gum and Vela clumps. The extents of the Gum I, Gum II, and Vela clumps are indicated by the light yellow and orange circles, overlaid on an H$\alpha$ intensity map (grey scale; \citealt{2003ApJS..146..407F}). Red points indicate pulsars with scattering delays $\tau$ that are underestimated by the model based on DM, while blue points indicate pulsars with $\tau$ overestimated by the model. The relative sizes of the points show the magnitude of ${\rm log}_{10}(\tau_{\rm mod}/\tau_{\rm obs})$. NE2001 is shown on the left, and the refitted model is shown on the right. For NE2001, the mean and rms of ${\rm log}_{10}(\tau_{\rm mod}/\tau_{\rm obs})$ are $-1.1$ and $2.0$, respectively; they decrease to $-0.0003$ and $0.59$ for \model. }
    \label{fig:gum-vela}
\end{figure*}

NE2001 models this region as three clumps (overdensities) that partially overlap on the sky; these clumps are referred to as Gum I, Gum II, and Vela. Figure~\ref{fig:gum-vela} shows the locations of these clumps superposed on an H$\alpha$ intensity map of the region, as well as pulsars that probe the region via measured scattering. In the model, all three clumps induce significant DM (having internal densities $n_{e,\rm I} = 0.43$ cm$^{-3}$, $n_{e,\rm II} = 0.5$ cm$^{-3}$, and $n_{e,\rm Vela} = 4.0$ cm$^{-3}$), but only Gum II and Vela have $F_c>0$. The original model parameters were motivated both by a large electron density required to explain the observed DMs of several pulsars in that direction \citep{1993ApJ...411..674T}, but also to explain significant scattering of both J0835$-$4510 and nearby pulsars within a $\approx 8^\circ$ radius \citep{mitra2001}. More recent observations have continued to affirm Gum and Vela's influence on pulsar scattering and DM \citep{purcell2015,kirsten2019,askew2025,krishnakumar2025}, but suggest that the spatial extent of Gum's scattering influence is larger than in the original model \citep{krishnakumar15,he2025,wang2025}.

\subsection{Vela Supernova Remnant}

In NE2001, the Vela supernova remnant has a distance of 250 pc and a radius of 38 pc. The parallax distance to the Vela pulsar, J0835$-$4510, is $287\pm18$ pc \citep{dodson2003}; for this observed distance, NE2001 predicts $\rm DM_{NE2001} = 243.9$ pc cm$^{-3}$, significantly greater than the observed $\rm DM_{obs} = 67.7$ pc cm$^{-3}$. This discrepancy is due to the location of the Vela clump along the LOS -- in the model, the LOS encounters the edge of the Vela clump at a distance of 222 pc, and passes through the entirety of the Vela clump before reaching the pulsar's observed distance of 287 pc. Fixing this discrepancy only requires placing the Vela clump further along the LOS, so that the pulsar is embedded in the clump rather than beyond it. Indeed, placing the location of the Vela clump near the parallax distance of the pulsar, at $D_{\rm NE2025}^{\rm Vela} = 300$ pc, yields a revised DM prediction of $68.1$ pc cm$^{-3}$, close to the observed value. This modification brings the location of the Vela clump in line with other studies, which typically use the Vela pulsar's distance as a proxy for the distance to the remnant (\citealt{cha99,chen2025_snr}; see, however, \citealt{gao2025} for a parsec-resolution 3D dust map of the Vela shell that places its center closer to 350 pc).

Increasing the distance to the Vela clump in the model reduces its effective angular size, causing some pulsars to no longer intersect the clump in the model. Given that these pulsars require the presence of the clump to explain their observed scattering, we increase the radius of the Vela clump to 44 pc, so that the angular size of the clump remains the same between NE2001 and the updated model. 

\subsection{Gum Nebula}

In Section~\ref{sec:model-comparison}, we identified three pulsars with $\tau$ severely underestimated by the model, both of which are viewed through Gum I. The lefthand panel of Figure~\ref{fig:gum-vela} illustrates that this discrepancy applies, in fact, to most of the pulsars with scattering measurements viewed through Gum I, suggesting that an increase in the fluctuation parameter $F_c$ of the clump is required. Given the spatial overlap between the Gum I, Gum II, and Vela clumps, we fit $F_c$ for all three  in tandem. Using 17 pulsars with scattering measurements viewed through the region, we optimize $(F_{c,\rm I}, F_{c,\rm II}, F_{c,\rm Vela})$ by minimizing both the mean and standard deviation of ${\rm log_{10}}(\tau_{\rm mod}/\tau_{\rm obs})$. We find that the new optimal values are $F_{c,\rm I} = 0.071$, $F_{c,\rm II} = 1.2$, and $F_{c,\rm Vela} = 1.9$, which yield a mean and rms of ${\rm log_{10}}(\tau_{\rm mod}/\tau_{\rm obs})$ of $-0.0003$ and $0.59$, respectively. These values are found after modifying the thick disk and spiral arms based on the results in Sections~\ref{sec:thick}--~\ref{sec:spiral} (although those other components have a minimal effect on the nearby, low-latitude sightlines through the Gum and Vela clumps). For comparison, the original NE2001 parameters yield a mean and rms of ${\rm log_{10}}(\tau_{\rm mod}/\tau_{\rm obs})$ of $-1.1$ and $2.0$, respectively, indicating that the revised parameters yield over an order of magnitude improvement in the scattering predictions. \revision{The revised parameters for both the Gum and Vela clumps are shown in Appendix Table~\ref{tab:clumps}.}

While the Gum and Vela clumps dominate scattering predictions in this direction, there is mixed observational evidence for whether they act as dominant scattering screens for all pulsars in the region. \cite{he2025} find evidence for enhanced scattering through the Vela clump based on a broad assessment of pulsars with observed scattering near and surrounding the remnant, \revision{although \cite{kumar2026} suggest the scattering is in fact dominated by the front edge of the Gum nebula}. Measurements of interstellar scintillations yield accurate fractional locations of discrete scattering regions along a pulsar LOS from fitting to scintillation arcs \citep[e.g.][]{stinebring2001} and using a precise pulsar distance.   Associating these screen locations with particular macroscopic ISM structures is often ambiguous, however. For example, while assessment of scintillation arcs for B0740$-$28 indicate a scattering screen at the edge of the Gum Nebula \citep{wang2025}, \cite{xu2023_2} find that scintillation of the Vela pulsar is more consistent with a scattering screen at the edge of the Local Bubble. Future models would benefit greatly from dedicated scintillation studies of as many pulsars in the Gum and Vela regions as possible, in order to determine whether they consistently act as the dominant scattering screens along the LOS. 

\section{Galactic Center}\label{sec:GC}

\begin{figure*}
    \centering
    \includegraphics[width=\textwidth]{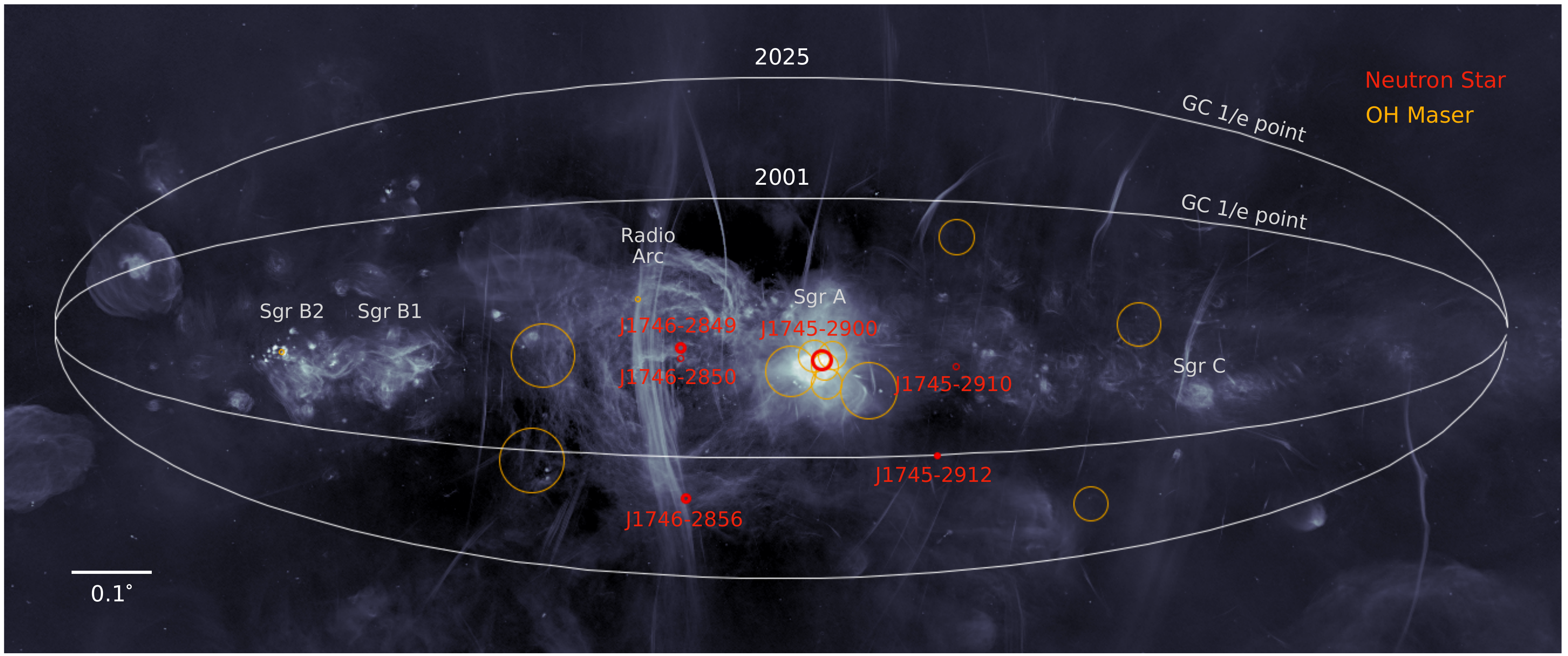}
    \caption{Distribution of neutron stars (red) and OH masers (orange) with measured scattering, within a $\approx 1^\circ \times 0.5^\circ$ region around the Galactic Center (GC), projected onto the MeerKAT radio image from \cite{heywood22}. The sizes of the red and orange circles indicate the relative amounts of observed angular broadening at 1 GHz (size $\propto \theta_{\rm d}$). The circles do not reflect the anisotropy that is observed in most of these sources' scattered images; here we show the geometric mean of the major and minor axes of the scattered image, scaled to 1 GHz assuming $\theta_{\rm d}\propto \nu^{-2}$. Two pulsars with DM-based distances within the GC region, but without measured scattering, are shown by small red circles (J1746$-$2850 and J1745$-$2910). The $1/e$ points of the GC components in NE2001 and \model\ are shown by the large white ellipses. The image is in Galactic coordinates with North up and East left. Several known structures in the GC are noted.}
    \label{fig:GC-meerkat}
\end{figure*}

The only region where NE2001 severely \textit{over}estimates scattering is the Galactic Center (GC). The GC component takes the form of an oblate spheroid:
\begin{equation}
    n_{\rm GC}(r_\perp,z) = n_{\rm GC0}e^{-[\delta r_\perp^2/R_{\rm GC}^2 + (z-z_{\rm GC})^2/H_{\rm GC}^2]},
\end{equation}
where $\delta r_\perp^2 = (x-x_{\rm GC})^2 + (y-y_{\rm GC})^2$. The original model parameters $(n_{\rm GC0},R_{\rm GC},H_{\rm GC}) = (10.0\ {\rm cm^{-3}}, 0.145\ {\rm kpc}, 0.026\ {\rm kpc})$ and $F_c = 0.6\times10^5$ were based primarily on angular broadening measurements of Sgr A* and OH masers, which suggested the presence of a hyperstrong scattering region spatially extended up to $\approx 0.5^\circ$ around Sgr A* \citep{langevelde91,langevelde92,frail94,lazio98}, spatially coincident with the Central Molecular Zone, a region of dense molecular gas $\approx 200$ pc in radius around the GC \citep{morris96,battersby25}. A key limitation of the OH masers is that they lack distance measurements, and the attribution of their scattering to the GC region is based on both their observed IR fluxes (which combined with an assumed stellar luminosity yields an estimated distance), as well as the similar magnitudes (and anisotropy) of their observed $\theta_{\rm d}$ and that of Sgr A* \citep{langevelde92}.

Radio detection of the GC magnetar J1745$-$2900 \citep{eatough13} led to  the first measurement of pulse broadening within $3^{\prime\prime}$ of Sgr A*, revealing a scattering delay $10^3$ times smaller than predicted by NE2001 \citep{spitler14_gc}. The  near equality of  the scattering diameters ($\theta_{\rm d}$) of the GC magnetar and Sgr A* suggests that they are observed through  the same scattering medium.  Assuming the medium comprises  a single thin screen implies a `far' distance   $\approx 6$ kpc from the GC \citep{bower14} rather than the `near' medium in the NE2001 model.   This far distance corresponds to a  high probability  traversal of  at least one \HII region in a foreground spiral arm \citep{sicheneder17}. 

Although scattering from a region near the GC (e.g. $\lesssim 1$\,kpc) is far less than in the NE2001 model, it is by no means nonexistent. Three other heavily scattered pulsars within $0.2^\circ$ of Sgr A* have observed $\theta_{\rm d}$ that are much smaller than that of the GC magnetar, despite having comparable $\tau$ -- all three of these pulsars have implied screen distances consistent with scattering in the vicinity of the GC \citep{dexter17}.  In a follow-on analysis of  the angular and pulse broadening measurements of these pulsars, \citet[][]{dexter17} estimated GC-screen distances for J1745$-$2912, J1746$-$2849, and J1746$-$2856
of $\dsl = 0.3\pm0.3, 0.9\pm0.7$ and $1.4\pm0.3$\,kpc, respectively.  

All of the scattering screen distances referred to above suffer from systematic uncertainties. As with analyses of the combined scattering of Sgr A* and J1745$-$2900, \citet[][]{dexter17}  extrapolated pulse broadening times measured at lower frequencies to the frequencies of the VLBI measurements of angular broadening. This approach potentially introduces a systematic error in estimated screen distances owing to the strong frequency dependence of $\tau$ and uncertainties in its scaling index with frequency.  Angular broadening measurements were found to scale as $\nu^{-2}$, as expected from a medium where the diffraction/scattering is dominated by the smallest length scale in electron density variations.  However,  pulse-shape modeling yielded scattering times $\tau \propto \nu^{-\xtau}$  with  $\xtau < 4$, more shallowly than the $\xtau=4$ value expected from the angular broadening. By scaling $\tau$ to higher frequencies rather than scaling $\thetad$ to lower frequencies, the far screen distance can be overestimated by as much as 1\,kpc. Moreover, usage of a Gaussian function as a model for the emitted pulse shape combined with an exponential PBF typically induces frequency-dependent biases into estimates of $\tau$ and hence in $\xtau$ \citep{geiger25}.  If the actual index for J1745$-$2900 is $\xtau=4$,  the published GC-screen distances for the single-screen model are in fact biased low because $\tau$ values extrapolated to the higher VLBI frequencies are biased high.  

Another revision of the far screen distance arises from the likely presence of multiple scattering screens along the LOS. With two screens contributing to scattering and taking into account geometrical leveraging, it can be shown  that appreciable scattering in the `near'  screen can cause the estimated distance of the `far' screen to be larger than in the single screen configuration. The net effect of the issues described above is that the far screen's actual distance from the GC is fairly uncertain, ranging from $\sim4$ to 7\,kpc.   This screen location can be compared to distances from the GC  to the centroids of the Crux-Scutum and Carina-Sagittarius spiral arms at $\sim 3.5$ and 7\,kpc and a distance of 3.8\,kpc for the peak of the annular thin disk.\footnote{Note these distances are based on a GC-Sun distance of 8.5\,kpc that is built into the NE2001 model.  A future model will use a slightly smaller distance.}

While a detailed model for GC scattering is beyond the scope of this update to NE2001, a plausible configuration has  two screen-like regions   for the  direction of the GC combined with large-scale scattering from spiral arms and disks as with any other LOS to the inner Galaxy.   In addition to Sgr A*, the magnetar, and nearby pulsars, strong scattering is also seen from OH masers within about $0.5^\circ$ of Sgr A*, which have considerable variability in their amount of scattering, possibly due to differing distances of the maser sources behind a scattering screen in the GC.   

Figure~\ref{fig:GC-meerkat} shows the sky distribution of the pulsars, the GC magnetar, and the OH masers, along with the relative magnitudes of their angular broadening, projected onto a MeerKAT radio image of the GC. The distribution of both $\tau$ and $\theta_{\rm d}$ supports previous inference that the GC scattering medium is non-uniform on scales as small as $\approx 0.1^\circ$ \citep{lazio99}, broadly consistent with a wealth of independent observations indicating an inhomogeneous, turbulent and multi-phase gas distribution in the GC \citep[][]{morris96,zhao10,ginsburg16,bally25}.  Pulsars close to the GC show large spatial variations in RM across the GC region by magnitudes up to $\approx 3000$ rad m$^{-2}$ \citep{abbate2023}.   These variations along with the 
radio image in Figure~\ref{fig:GC-meerkat} illustrate the  severe complexity of the gas and magnetic field distribution in the region.

\begin{figure*}
    \centering
    \includegraphics[width=0.9\textwidth]{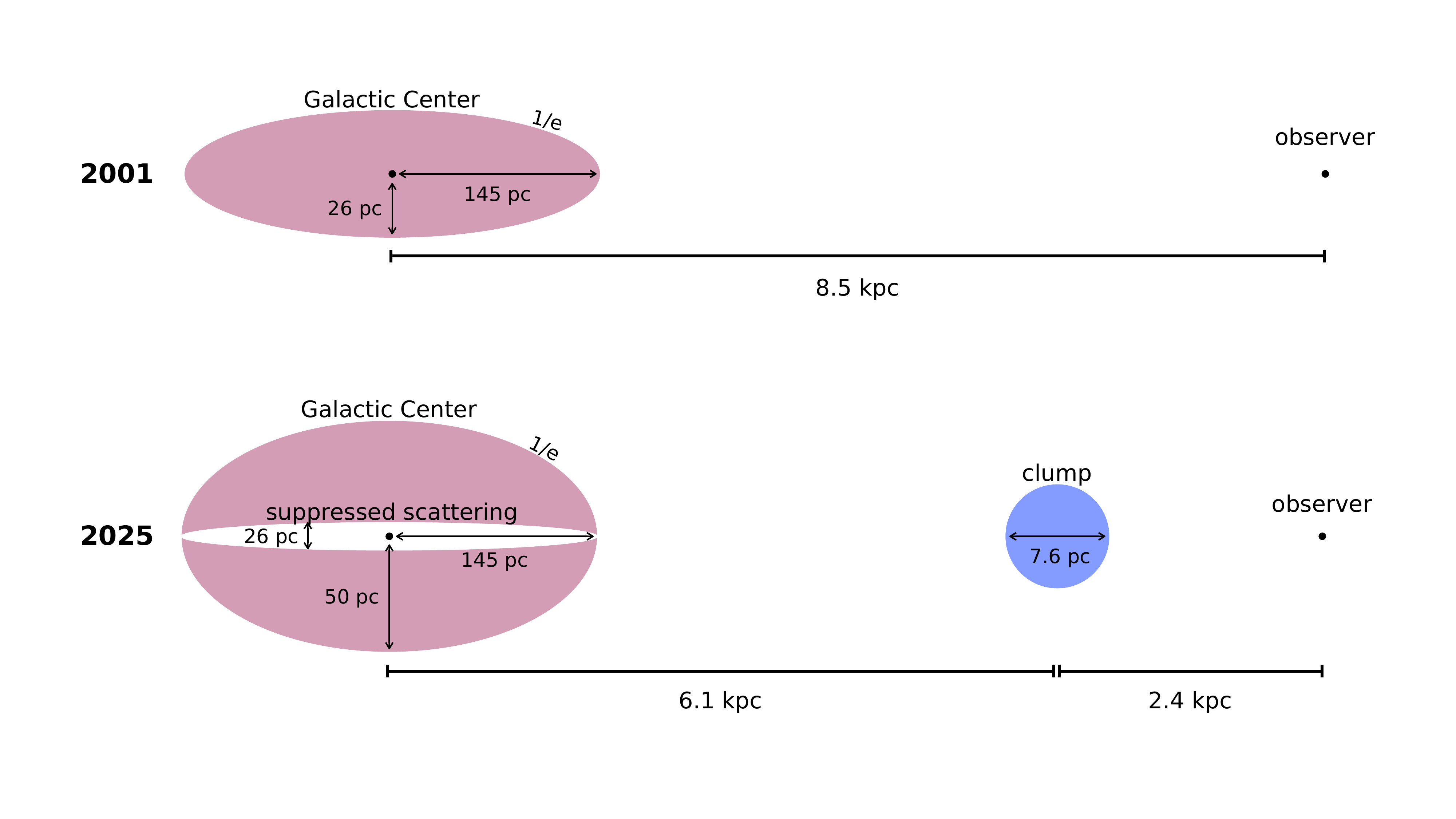}
    \caption{Schematic illustrating the geometries of the Galactic Center (GC) region used in NE2001 and the updated model (not to scale). For the GC component (pink), the scale height and radius indicate the $1/e$ extent of the electron density, while the clump added to the model (blue) has the cutoff radius indicated in the diagram. A region of suppressed scattering (white) is introduced in the updated model to account for the reduced scattering strength of the GC around Sgr A* and the GC magnetar J1745$-$2900 (see Section~\ref{sec:GC}).}
    \label{fig:gc-diagram}
\end{figure*}

Nonetheless, a conservative GC model should remain as simple as possible, given the small number of pulsars that trace the region and also that we are presenting a minimalist update to the NE2001 model. We therefore take the following approach: The $F_c$ parameter of the GC component is reduced significantly, in order to achieve consistency with the three GC pulsars that are observationally indicative of scattering from the GC (excluding J1745$-$2900). Due to the limited number of data points, we fit for $F_c$ primarily by minimizing the absolute mean logarithmic difference between predicted and observed scattering measures. Two of the pulsars are actually outside $H_{\rm GC}$ (the $1/e$ height) of the original GC component, and so we expand the GC component in the $z$-direction to encompass these pulsars (see Figure~\ref{fig:GC-meerkat}). For the GC magnetar and Sgr A*, we introduce a suppression in the scattering strength of the GC in a small ($0.09^\circ$ radius) sphere around the magnetar. Given evidence for foreground scattering of the GC magnetar and Sgr A*, a clump is added to the model with a distance and $F_c$ determined through simultaneous fitting of the magnetar's observed $\tau$ and $\theta_{\rm d}$; the clump density is left small in order to produce minimal contribution to DM and hence have minimal effect on the predicted distance. The \revision{solution for the} clump distance of 6.1 kpc from the GC \revision{places it at the edge of the Car-Sgr arm in the model, and} is within the broad range of values we consider plausible based on the discussion of screen distances above \revision{(broadly consistent with far screen distances inferred by \citealt{bower14} and \citealt{dexter17})}. Figure~\ref{fig:gc-diagram} schematically illustrates the change in the scattering geometry between NE2001 and the updated model.

Table~\ref{tab:gc} shows how the observed scattering of the four GC neutron stars compares to both the NE2001 and \model\ predictions. The pulse broadening predictions are substantially improved in the updated model, and they are on average $10^3$ times smaller than NE2001; i.e., on average they completely close the gap between the observed $\tau$ and the original model predictions. Unfortunately, the $\theta_{\rm d}$ predictions are not improved, and in the updated model they are systematically underestimated. Nonetheless, the average difference between the observed and predicted $\theta_{\rm d}$ is still within a factor of 2, within the error tolerance of the original model. Fully reconciling the observed and predicted $\tau$ and $\theta_{\rm d}$ would require placing the scattering screens for J1746$-$2849 and J1746$-$2856 further away from the GC, adding additional complexity to the model (and potentially leading to over-fitting given the small number of data points). 

\begin{deluxetable*}{c C C C C C C C C C}\label{tab:gc}
\tabletypesize{\scriptsize}
\tablecaption{Scattering Predictions for Galactic Center Neutron Stars}
\tablehead{\colhead{} & \colhead{} & \colhead{} & \colhead{} & \colhead{} & \colhead{} & \multicolumn{2}{c}{{2001}} & \multicolumn{2}{c}{{2025}} \\ \hline
\colhead{Name} & \colhead{$l$ $(^{\circ})$} & \colhead{$b$ $(^{\circ})$} & \colhead{DM (pc cm$^{-3}$)} & \colhead{$\tau_{\rm obs}$ (ms)} & \colhead{$\theta_{\rm d,obs}$ (mas)} & \colhead{$\tau_{\rm 2001}$ (ms)} & \colhead{$\theta_{\rm d,2001}$ (mas)} & \colhead{$\tau_{\rm 2025}$ (ms)} & \colhead{$\theta_{\rm d,2025}$ (mas)}}
\startdata
J1745$-$2900 & -0.056 & -0.047 & 1778.0 & 1300.0 & 946.0 & 3.75\times10^6 & 1732.0 & 1288.0 & 937.3 \\
J1745$-$2912 & -0.212 & -0.175 & 1130.0 & 3630.0 & 129.0 & 4.57\times10^5 & 392.0 & 1073.0 & 125.8 \\
J1746$-$2849 & 0.134 & -0.03 & 1456.0 & 1584.0 & 401.0 & 1.65\times10^6 & 944.0 & 3025.0 & 136.4 \\
J1746$-$2856 & 0.126 & -0.233 & 1168.0 & 747.0 & 379.0 & 5.70\times10^5 & 438.0 & 1296.0 & 126.0 \\
\hline
\multicolumn{6}{c}{Mean log$_{10}$ Difference from Observed} & 2.9 & 0.29 & -0.003 & -0.24 \\
\multicolumn{6}{c}{RMS log$_{10}$ Difference from Observed} & 0.49 & 0.16 & 0.32 & 0.23 \\
\enddata
\tablecomments{Left to right: Pulsar name, Galactic longitude and latitude, DM, observed scattering delay $\tau$ and angular broadening $\theta_{\rm d}$, and predicted $\tau$ and $\theta_{\rm d}$ from NE2001 and \model. All scattering is referenced to 1 GHz, assuming $\tau \propto \nu^{-4}$ and $\theta_{\rm d} \propto \nu^{-2}$. The $\theta_{\rm d}$ measurements are from \cite{dexter17}; the $\tau$ measurements are from \cite{johnston2006}, \cite{deneva2009}, and \cite{spitler14_gc}. The bottom two rows show the mean and rms of $\rm log_{\rm 10}(\tau_{mod}/\tau_{obs})$ and $\rm log_{\rm 10}(\theta_{d,mod}/\theta_{d,obs})$ for the four pulsars.}
\end{deluxetable*}

The revised GC model plausibly reflects the evidence for at least two scattering components toward the GC, one of which is much stronger than the other. It also captures the presence of strong foreground scattering in the direction of Sgr A*. However, several shortcomings remain. It is unclear whether the suppression of GC scattering along the LOS to J1745$-$2900 is entirely physically plausible. While Sgr A* could potentially evacuate a cavity in its surrounding plasma, it is unclear whether such a process would dampen density fluctuations. 
Another shortcoming of the revised model is that it no longer reproduces the observed $\theta_{\rm d}$ of OH masers at larger angular separations from Sgr A* (i.e., the data used to calibrate NE2001), if one assumes that the OH masers are about 8.5 kpc away and are scattered by the GC (as was assumed in NE2001). Several factors could account for this discrepancy, including the poorly constrained maser distances, as well as patchiness in the scattering screen(s). Most of the OH maser $\theta_{\rm d}$ measurements are also based on a single observing frequency, and hence lack empirical constraints on the frequency scaling of scattering along the LOS. 

Ultimately, a future model should explore a completely new analytic description of the GC that reflects the complicated picture painted above. An alternative approach could reconcile the OH maser and neutron star scattering measurements by leaving the fluctuation parameter of the GC region large, and instead introduce voids in front of each neutron star. While this approach is less desirable due to the larger number of free parameters, the continued low discovery rate of radio pulsars in the GC suggests that the small scattering delays of currently known GC pulsars could be exceptions rather than indicative of the GC's general scattering properties. We also note that while we have continued to adopt the GC distance used in NE2001, 8.5 kpc, due to its covariance with many of the other model components, it is possible that a number of the issues discussed above will be modified slightly in a future model by the use of a smaller distance that is in line with more recent observations \citep[e.g.,][]{reid2019}. 

\section{Clumps and Voids}\label{sec:clumps}

The contributions to pulsar DMs from \HII\ regions along the LOS were recognized by \citet[][]{prentice-terhaar1969}, and ISM underdensities corresponding to supernovae-driven bubbles and Galactic chimneys have also long been thought relevant to pulsars with anomalously low DMs, as well as enhanced scattering \citep[e.g.][]{1999ApJ...523L.171T,bhat2002}. Clump and void features were thus incorporated in the NE2001 model, and the same is true for \model. While the majority of pulsar measurements considered in this study are adequately fit by smooth large-scale structures with homogeneous scattering properties (thin and thick disks, spiral arms), there remain a minority of pulsars that have DM and scattering measures that are significantly discrepant from model predictions. These outlier pulsars indicate the presence of smaller-scale ISM structures, as evidenced, e.g., by the identification of \HII regions along a number of these outlier pulsars' sightlines (see Section~\ref{sec:model-comparison}; \citealt{ocker2024c}). 

A more minimalist approach adopted by \cite{ymw16} and some previous models \citep{cordes91,1993ApJ...411..674T,gomez2001} is to ignore these outliers and only fit for the Galaxy's large-scale structure, while including local ISM features like the Local Bubble and Gum Nebula on a model-dependent basis. Such an approach ignores robust observational evidence that small-scale ISM structures can dominate the observed properties of some pulsars and quasars \citep{spangler1998,rickett2009,mall2022}; indeed, small-scale overdensities like \HII regions appear to be increasingly common along pulsar sightlines as the known pulsar population expands \citep{ocker2024c}. Properly accounting for these inhomogeneities increases model accuracy and provides additional constraining power for investigating known ISM structures, as we will show in due course. Nonetheless, \model\ users can opt to exclude clumps and voids when evaluating specific sightlines, and diagnostic tools are provided with the model to test clump and void contributions. 

Clumps and voids are included in the model based on three main lines of evidence (assessed prior to clump/void inclusion): (1) The distance, DM, and/or scattering of a pulsar is significantly misestimated; (2) the observed DM is at least $10$ pc cm$^{-3}$ greater than the maximum model prediction integrated through the entire disk along that sightline;\footnote{We consider 10 pc cm$^{-3}$ a conservative lower bound on the DM contribution of the Galactic halo \citep{2020MNRAS.496L.106K}. Pulsars with DMs more than 10 pc cm$^{-3}$ greater than the model maximum would be implausibly placed by the model either within the halo or beyond the Galaxy entirely.} and (3) there is a significant gap in the DM distribution of pulsars within a uniformly surveyed sky area (see Section~\ref{sec:gum55}). Outliers in distance and DM are taken to be pulsars in the 90th percentile of $|{\rm log_{10}}(D_{\rm obs}/D_{\rm mod})|$ and $|\rm log_{10}(DM_{\rm obs}/DM_{\rm mod})|$, respectively. Outliers in scattering are identified using a slightly stricter criterion, the 95th percentile in $|{\rm log_{10}}(\tau_{\rm obs}/\tau_{\rm mod})|$. Overestimated distances (underestimated DM and scattering) are used to constrain clumps, and underestimated distances (overestimated DM and scattering) are used to constrain voids. Some pre-existing clumps and voids from NE2001 are either removed or modified. Some clumps and voids are fit using multiple pulsar sightlines, such as the Cygnus region and the Fermi Bubbles. An overview of the clumps and voids included in the model is given below, followed by \revision{an assessment of their impact on the model and} an in-depth treatment for specific regions of interest. Appendix Tables~\ref{tab:clumps} and \ref{tab:voids} list the clumps and voids that are new or revised in \model.

\textit{Overview of Clumps:} \revision{There are 214 clumps in \model.} In total, 73 clumps have been added to the pre-existing clumps from NE2001, while 32 clumps from NE2001 have been removed. Fourteen of the new clumps are required to mitigate significant overestimation of pulsar distances; 41 clumps are required to mitigate pulsars with DMs in excess of the model's maximum prediction; and 28 clumps mitigate significant underestimation of pulsar scattering. One clump is added to explain a significant gap in the DM distribution of pulsars between $-20^\circ\lesssim l\lesssim -10^\circ$ (the hole identified in Figure~\ref{fig:dm-vs-lon1}; see Section~\ref{sec:gum55}). Thirty of the clumps added to the model correspond to known \HII regions, the distances and radii of which we take from the \cite{houhan2014} and \cite{wisecat_2014} \HII region catalogs where possible. Twenty-one of the clumps added have multiple sightlines through them. We consider clumps associated with \HII regions and/or multiple pulsars to be the most robust in the model; these clumps are noted in Table~\ref{tab:clumps} and in the model code. In the absence of a known \HII region or constraints from multiple sightlines, the clump distance is taken to be half the pulsar distance and the clump radius is taken to be 10 pc. Two of the largest regions modeled are Cygnus and the area around Gum 55, which are discussed further below. 

\textit{Overview of Voids:} \revision{There are 37 voids in \model.} In total, 22 voids have been added to the pre-existing voids from NE2001, two of which replace voids from the original model. Five voids are required to mitigate significant overestimation of scattering, while the rest are required due to significant underestimation of distance (in many cases a result of the increase in spiral arm density). Five of the voids added have multiple sightlines through them, including voids in the Car-Sgr and Perseus arms, as well as the Fermi Bubbles (see Section~\ref{sec:fermi}). Voids are generally placed in spiral arms, with a central density $\geq 0.001$ cm$^{-3}$ (comparable to the densities inferred in Galactic superbubbles; e.g. \citealt{1998ApJ...498..689H}). Similar to the clumps, voids with multiple sightlines through them are likely the most robust, and are noted in Table~\ref{tab:voids}. Three globular clusters (NGC5986, M80, NGC6342) require multiple voids to resolve their deficit in observed DM from the model. One void placed in the Perseus arm at $l=129^\circ$ (a modification to an existing void in NE2001) reconciles the underestimation of two pulsar distances and the overestimation of the Galactic DM contribution for the extragalactic FRB 20220319D \citep{ravi2025}. The largest voids, the Fermi Bubbles, are discussed further below.

\subsection{Impact of Clumps and Voids on the Model}

\revision{There are many more clumps and voids known in the ISM than are included in \model, which is limited by the sparse distribution of known pulsars in the Galaxy. Nonetheless, the fraction of all known pulsar sightlines that intersect a clump or void in \model\ is not insubstantial: Out of the $\sim4100$ Galactic radio pulsars known, about 1800 have sightlines intersecting a clump in NE2025, corresponding to about 43\% of the known population. This is slightly larger than the expected fraction of pulsar sightlines through HII regions, which is estimated to be up to 33\% in \cite{ocker2024c}. The fraction of known pulsar sightlines that intersect voids in NE2025 is larger: 2123 pulsars, or about 50\%. However, most of these intersections are near the edge of a clump or void, and have little impact on the DM and scattering predictions compared to intersections within the $1/e$ radius. When we instead evaluate how many known pulsar sightlines come within the $1/e$ radius of a clump (i.e., cases where the clump can dominate the DM and/or scattering), the number is reduced dramatically to just 194 sightlines (5\%); the number increases to 704 sightlines (17\%) for intersections within the $2/e$ radius. A similar reduction in intersections applies to voids. These fractions are also highly dependent on Galactic latitude, as discussed further below. In summary, a small fraction of known pulsars have DM and scattering predictions that are dominated by clumps and voids in the model.}

\revision{Given that the pulsar population is not distributed uniformly throughout the ISM, but is rather heavily biased towards low Galactic latitudes where the chance of clump/void intersection is larger, it is also worth assessing the chance of intersecting a clump or void for any given direction on the sky. To that end, we have simulated $10^4$ randomly generated sightlines that are distributed uniformly in sky location and distance, for $10^\circ$-wide latitude bins. The chance of intersecting a clump declines from $31\%$ at $|b|<10^\circ$ to $\lesssim8\%$ at $|b|\geq 30^\circ$, while the chance of intersecting a void goes from $27\%$ at $|b|<10^\circ$ to $\lesssim14\%$ at $|b|\geq30^\circ$. The void intersection probability declines more slowly with latitude due primarily to the Fermi Bubbles. For high latitudes ($|b|>20^\circ$), we find intersection probabilities comparable to the finding in \cite{ocker2020} that about $15\%$ of pulsar sightlines at $|b|>20^\circ$ intersect clumps and voids.}

\revision{The probability of intersecting a clump or void is strongly affected by the largest of these structures, which are described in detail below.}

\subsection{Cygnus}\label{sec:cygnus}

\begin{figure*}
    \centering
    \includegraphics[width=\linewidth]{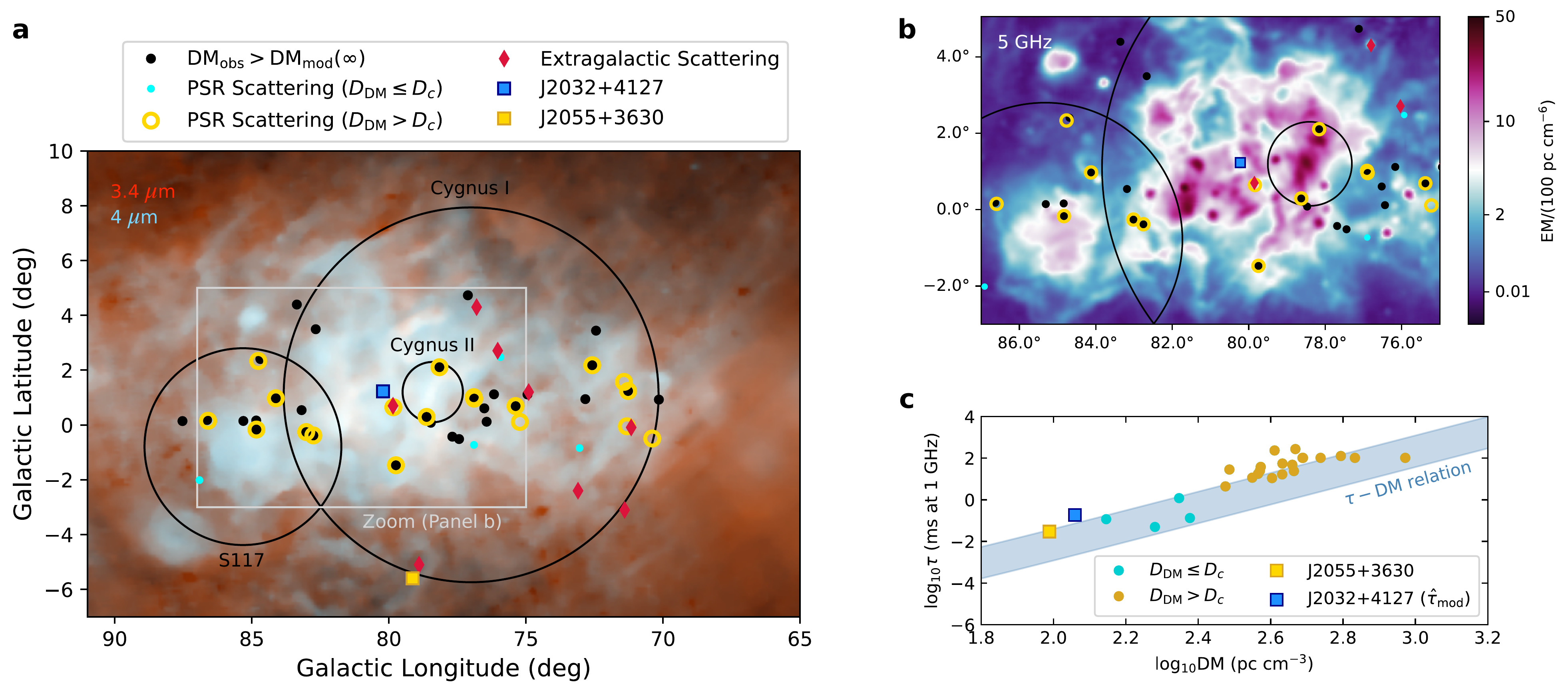}
    \caption{Schematic view of the Cygnus region as modeled in \model. (a) A cutout from the SPHEREx all-sky mosaic at 3.4 $\mu$m and 4 $\mu$m; figure credit: NASA/JPL-Caltech (\url{https://spherex.caltech.edu}; \citealt{{crill2020,bock2025}}). The three clumps included in the model, Cygnus I, II, and S117, are shown by the black circles. Pulsars with $\rm DM_{obs} > DM_{mod}(\infty)$ (prior to clump fitting) are shown in black. Pulsars with scattering measurements used to fit the clumps are shown as yellow and cyan circles, which correspond to pulsars with model-based distances beyond Cygnus ($D_{\rm DM}>D_c$) and within Cygnus ($D_{\rm DM}\leq D_c$), respectively. Extragalactic sightlines (quasars and FRB20210705) with measured scattering are shown as red diamonds \citep{fey89,fey91,molnar95,zhou2023,patil2025}. Two pulsars with parallax distances that constrain the region are shown by the blue and yellow squares. A zoom-in on the grey rectangle is shown in panel (b), which is a map of emission measures inferred from total intensity at 5 GHz measured by the Sino-German 6 cm Survey (\citealt{gao2019} and refs. therein; see \citealt{ocker2024c} for details). (c) Scattering time $\tau$ vs. DM for the same pulsars shown in panels (a) and (b). J2032$+$4127 does not have a published scattering time, and the model-predicted $\tau$ is shown instead. The $\pm1\sigma$ contours of the $\tau$-DM relation \citep{cordes2022} are shown by the shaded blue region.}
    \label{fig:cygnus}
\end{figure*}

The Cygnus region is the largest clump in angular size that is new in \model. 
NE2001 included 16 clumps in this direction, intended to model the significant angular broadening of background quasars \citep{spangler1998,lazio1990}; at the time, no known pulsars could be used to place robust constraints on the DM and scattering contribution of the region. The FAST Galactic Plane Pulsar Survey has since discovered several pulsars in the direction of Cygnus with observed DMs greater than the maximum DM predicted by both NE2001 and YMW16 \citep{han2021_fast,han2025}, indicating that the region has a substantial electron column density \citep{ocker2024c}. New constraints on the electron density distribution of Cygnus are also enabled by a detection gap in the distribution of FRBs observed by CHIME, suggesting that Cygnus produces sufficient amounts of scattering ($\tau \gtrsim 5.6$ ms at 1 GHz) to suppress the FRB detection rate \citep{patil2025}. As shown by \cite{patil2025}, observations of pulsars, FRBs, and quasars viewed in this direction are all consistent with Cygnus being the dominant scattering medium for these background sources.

We replace the 16 clumps of NE2001 in this direction with two large clumps, Cygnus I and S117, as well as one smaller clump (Cygnus II). Figure~\ref{fig:cygnus} shows a schematic view of the region. The central electron density and radius of the Cygnus I clump are bound by two pulsars with parallax distances: J2032$+$4127 ($l=80.224^\circ, b=1.028^\circ, D_{\rm kpc}=1.45\pm0.07$) and J2055$+$3630 ($l=79.133^\circ,b=-5.589^\circ,D_{\rm kpc}=5.9^{+1.3}_{-0.9}$). Without including Cygnus I, the model underestimates the DM of J2032$+$4127 by 98 pc cm$^{-3}$, while the DM of J2055$+$3630 is overestimated by 13 pc cm$^{-3}$. We use the DM deficit in the model for J2032$+$4127, combined with an adopted distance to Cygnus I of 1.5 kpc \citep{rygl2012}, central coordinates $(l=77.7^\circ,b=1.1^\circ)$, and the requirement that the clump not overlap with the J2055$+$3630 LOS, to find a peak central density of $n_e = 0.9$ cm$^{-3}$ and a radius of $145$ pc. The Cygnus I clump alone is insufficient to explain the large DMs of pulsars within a part of the Cygnus complex between $78^\circ\lesssim l \lesssim 87^\circ$. About ten pulsars in this range are seen through and around the \HII region S117, which is modeled as a separate clump with a central density and radius set to minimize $\rm DM_{\rm obs} - DM_{\rm mod}(\infty)$ for those pulsars. An additional two pulsars within the Cygnus I clump still have observed DMs greater than the model maximum, even after including Cygnus I and S117. These pulsars coincide on the sky with a section of enhanced radio emission within the broader Cygnus complex (see panel (b) in Figure~\ref{fig:cygnus}). A second clump, Cygnus II, is therefore introduced as a smaller clump embedded within Cygnus I, with a central density and radius set to make the model's maximum DM greater than the observed DM for these pulsars.

The scattering strength of Cygnus, parameterized by $F_c$, is fit using 23 pulsars with scattering measurements in this direction (Figure~\ref{fig:cygnus}c). Here we also find evidence for spatial variations in the scattering strength of Cygnus, as the final best-fit values for Cygnus I, II, and S117 are $F_{c,\rm I} = 4.7$, $F_{c,\rm II} = 1.8$, and $F_{c,\rm S117} = 10$. The peak scattering times $\tau$ caused by each clump are more than sufficient to explain the FRB detection gap reported by \cite{patil2025}: through Cygnus I, $\tau$ peaks at about 180 ms at 1 GHz for an extragalactic source, and through S117, $\tau$ is about $2\times$ smaller. A detailed assessment of FRB and quasar scattering measurements in this direction, and how they compare to \model, will be considered in future work. 

At the edge of Cygnus I, J2055$+$3630 receives little DM contribution from the clump. In fact, the DM of this pulsar is so small for its parallax distance that it meets the criterion for including a void along the LOS (see Table~\ref{tab:voids}). To make matters more complex, the scattering time is larger than the model predicts for this pulsar's distance and DM, thereby also meeting the criterion for including a clump, but where that clump should be placed along the LOS is unclear (it could be a density enhancement at the edge of Cygnus, or elsewhere along the pulsar's 5.9 kpc-long sightline). While in the present model we include a small additional clump at the edge of Cygnus, thereby reproducing the observed scattering, this pulsar exemplifies the challenges of attempting to model substructure in the ISM, and thorough examination of this pulsar's scattering properties (e.g., targeted observations that constrain the scattering screen location) is needed.

\subsection{Gum 55 and the Near-Side of the Sagittarius Arm}\label{sec:gum55}
\begin{figure}
    \centering
    \includegraphics[width=0.45\textwidth]{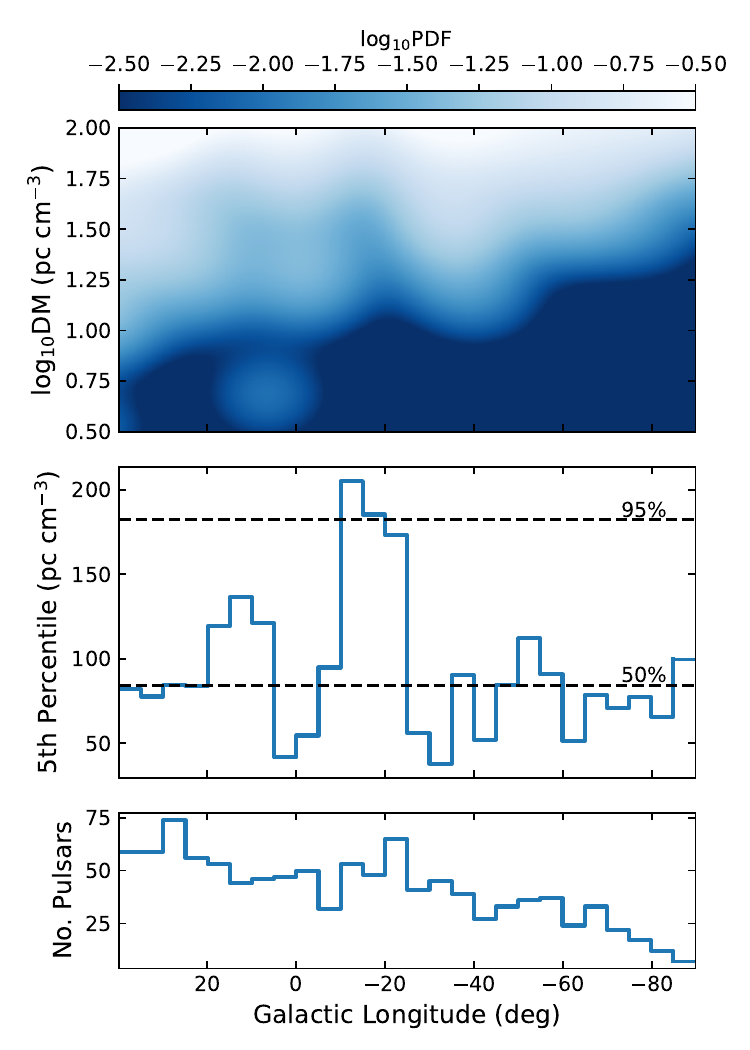}
    \caption{Statistical assessment of the gap in minimum DMs detected by the Parkes Multi-Beam Survey, corresponding to the ``hole'' identified in Figure~\ref{fig:dm-vs-lon1}. Top panel: Zoom-in on the PDF of DM shown in Figure~\ref{fig:dm-vs-lon1}. The ``hole'' is readily apparent as a decrease in probability at $\rm DM \gtrsim 100$ pc cm$^{-3}$ for $l \approx -20^\circ$. Middle panel: Fifth percentile of DM vs. Galactic longitude, calculated in $5^\circ$-wide longitude bins for $|b|<5^\circ$, within the uniform survey footprint. Black dashed lines indicate $50\%$ and $95\%$ confidence intervals of the minimum DM statistic. Bottom panel: Number of pulsars per bin vs. Galactic longitude. The longitude range shown has $N\geq 20$ pulsars per bin, ensuring sufficient sample sizes for evaluating the 5th percentile. The drop-off in pulsars detected at $l < -60^\circ$ may be responsible for the differing trend between the DM PDF (which is based on a Gaussian kernel density estimator evaluated in logarithmically-spaced DM bins) and the 5th percentile at these low longitudes.}
    \label{fig:hole-hist}
\end{figure}

In Section~\ref{sec:model-comparison}, we noted the apparent gap in pulsars with $\rm DM \lesssim 200$ pc cm$^{-3}$ around $l\approx -20^\circ$ (the ``hole'' in Figure~\ref{fig:dm-vs-lon1}). This gap was pointed out by \cite{ne20011}, but not explicitly modeled in NE2001. Such a gap is surprising because pulsar surveys tend to be most complete for low-DM (i.e., nearby) sources. To determine whether this gap is significant, we statistically evaluate the minimum DM of pulsars detected by the Parkes Multi-Beam Survey (PMBS) which has uniform coverage and sensitivity across the survey footprint ($50 \geq l \geq -100^\circ$, $|b|<5^\circ$) \citep{manchester2001,lorimer2006}. We focus solely on the observed DM distribution because it is entirely empirical, and we do not make any assumptions about the underlying pulsar distance or luminosity distributions. The most relevant selection effect biasing the DM distribution is scattering, which tends to suppress the detection of high-DM sources, not the low-DM sources of interest here. We define the minimum detected DM as the 5th percentile of the PMBS DM distribution, which is calculated within $5^\circ$-wide bins in longitude and 1 latitude bin across the survey footprint; the bin sizes are chosen to have a sufficient number of pulsars ($N\gtrsim 20$) within each bin to robustly calculate the 5th percentile. Figure~\ref{fig:hole-hist} shows the minimum DM as a function of longitude, as well as the number of pulsars within each longitude bin and a zoom-in on the DM PDF from Figure~\ref{fig:dm-vs-lon1}.

Significant outliers (above the 95th percentile) in the minimum DM are found between $-20^\circ \leq l \leq -10^\circ$: in this longitude range, the minimum DM is 190 pc cm$^{-3}$, nearly 100 pc cm$^{-3}$ larger than the median (84 pc cm$^{-3}$) across all longitude bins (translating to an $\approx1.3$ kpc difference in the minimum DM-based distance). The amount of enhancement in the minimum DM, as well as its exact longitude range, is sensitive to the spatial binning; however, finer binning leads to large variations in the number of pulsars per bin and insufficient sampling for robust statistics. As shown in Figure~\ref{fig:hole-hist}, the binning used here keeps the number of pulsars per bin roughly constant across the longitude range of interest, ensuring the statistics are approximately uniform. We also note that extending this procedure to pulsars beyond the PMBS footprint yields detection of a significant enhancement in the minimum DM around Cygnus, as well as other regions of enhanced density, affirming that there is a relation between the distribution of minimum DM and ISM structures.

There are several candidate ISM structures that may be responsible for the enhancement in minimum DM at this longitude range, including the \HII regions Gum 55, Gum 56, and Sh 2-2, which are spatially coincident with a region of larger H$\alpha$ nebulosity covering nearly $280$ square degrees. All three of these structures are located at similar distances ($D\approx 1.6 - 1.9$ kpc; \citealt{avedisova89,benaglia99,sung2013}), the closest being Gum 55 on the near side of the Sagittarius arm. The photometric distance to Gum 55, $D \approx 1.6$ kpc \citep{sung2013}, is just 200 pc above the distance predicted by \model\ for the median value of the minimum DM across the entire longitude range of PMBS.

Based on the evidence above, we include a clump at $(l=-15^\circ, b=1.1^\circ,D_{\rm kpc}=1.6)$ with a radius of 140 pc and a peak density $n_e = 0.5$ cm$^{-3}$ (giving a DM depth of 70\,pc\,cm$^{-3}$) that is calibrated such that the minimum DM-based distance at $-20^\circ \leq l \leq -10^\circ$ is consistent to $<2\sigma$ with the median across the entire PMBS footprint. The scattering strength of the clump is based primarily on the scattering time of J1705$-$3950, $\tau = 35\pm9$ ms at 610 MHz \citep{lewandowski2015}.

\subsection{Fermi Bubbles}\label{sec:fermi}

About a dozen globular clusters at $D_{\rm obs}>5$ kpc have distances significantly underestimated (by up to a factor of 5) in the model. These sightlines extend up to $|b|\lesssim 20^\circ$ and $|l|<50^\circ$. These same sightlines' distances are severely underestimated by NE2001 (see, e.g., Figure~\ref{fig:dist-diff}). The average DM excess in the model for these sightlines is 90 pc cm$^{-3}$; for a thick disk density $\leq 0.018$ cm$^{-3}$, this translates to a $>5$ kpc path length required through a void to recover the observed distances. We interpret this discrepancy between model and observations as plausible evidence for the Fermi Bubbles, large gamma-ray emitting cavities extending up to $|b|\approx 50^\circ$ and $|l|\approx 25^\circ$ that are thought to have been blown out by the Galactic Center \citep{su2010}. While the Fermi Bubbles were included in YMW16, only two sightlines were available to fit, and the Bubbles were ultimately left at the same density as the thick disk (i.e., having no impact; \citealt{ymw16}). 

The Fermi Bubbles are modeled as two triaxial ellipsoids above and below the Galactic Center. Due to the significant DM excess in the model, we assume a small, uniform internal density of $0.001$ cm$^{-3}$ and perform a grid search for the major and minor axes of the ellipsoids, seeking to minimize the $\chi^2$ difference between observed and modeled distances. We also test for different void centers, ultimately placing them at $l = 0^\circ, b = \pm 10^\circ$. The void dimensions are better constrained in the $z$-direction with a best-fit minor axis $ccv = 1.0\pm0.8$ kpc. The $x,y$ major axes are more poorly constrained; the $\chi^2$ is minimized for $(aav,bbv)=4.0$ kpc, with a $1\sigma$ upper limit of $5.0$ kpc and a lower limit that is essentially unbound. The final adopted values are thus $(aav,bbv,ccv) = (4.0,4.0,1.0)$ kpc. On the sky, the modeled Bubbles extend out to $l = \pm 28^\circ$ and $3^\circ < |b| < 22^\circ$, comparable in longitude range to the gamma-ray emission but smaller perpendicular to the plane \citep{su2010}. 

Four globular cluster sightlines through the Fermi Bubbles, between $-12.5^\circ \leq b \leq -7.9^\circ$, meet the criteria for an additional void along their LOSs. Given their close proximity on the sky, we model these sightlines using a single void in the Car-Sgr arm at $(l=2.21^\circ, b=-10.52^\circ,D_{\rm kpc}=1.5)$. 

\section{Discussion}\label{sec:conc}

We have used a sample of 171 pulsar distances, all with $<25\%$ fractional uncertainties, \revision{$4154$ DMs,} and 568 scattering measurements to update the NE2001 Galactic electron density model. Scattering measurements for FRBs, AGN, and masers have also informed the model. The thick disk, thin disk and spiral arms, Gum Nebula and Vela supernova remnant, and Galactic Center have been refitted, and a number of clumps and voids have been added to the model. The updated model, \model, provides significant improvement in distance and scattering estimation, as was shown in Figures~\ref{fig:dist-diff} and \ref{fig:tau-diff}. In the following subsections, we discuss the model performance in comparison to NE2001 and YMW16, consequences of the model for extragalactic studies, and directions for a future model.

\subsection{Model Performance and Comparison}

\begin{figure*}
    \centering
    \includegraphics[width=0.99\linewidth]{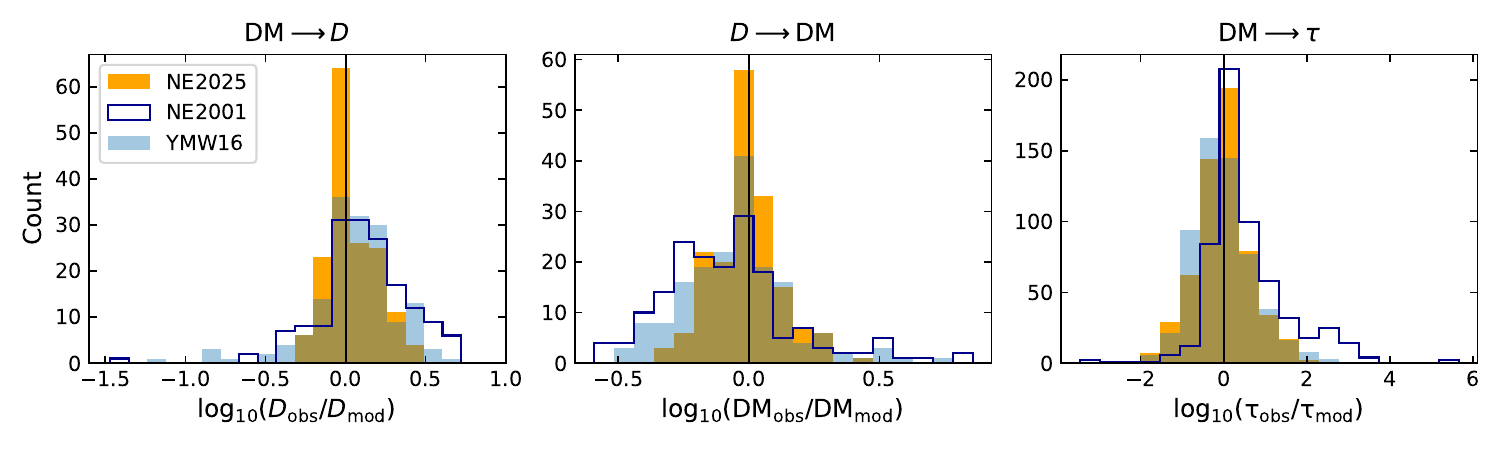}
    \caption{Logarithmic differences between observed and predicted distances (left), DMs (middle), and scattering times (right). Orange histograms show the differences for \model, unfilled histograms for NE2001, and light blue histograms for YMW16. As in earlier sections, the distances are predicted only for LOSs $<4$ kpc above the Galactic plane, whereas DMs are predicted for the entire distance sample.}
    \label{fig:error-hist}
\end{figure*}

Based on the data used in this work, the median fractional uncertainty in distance estimation $(D_{\rm obs}-D_{\rm mod})/D_{\rm obs}$ for \model\ is $0.9\%$ with an rms of $30\%$, compared to a median fractional uncertainty of $20\%$ with an rms of $200\%$ for NE2001, and a median fractional uncertainty of $15\%$ with an rms of $150\%$ for YMW16. The median log-difference in scattering estimation from DM $\rm log_{10}(\tau_{obs}/\tau_{mod})$ improves from $0.28$ in NE2001 to $-0.03$ in \model, with a reduction in rms from $0.98$ to $0.65$. Figure~\ref{fig:error-hist} shows the distribution of logarithmic differences between observed and predicted distances, DMs, and scattering times for \model, NE2001, and YMW16.

Among the most significant revisions in \model\ is the partitioning of free electrons between the thick disk and spiral arms. The updated thick disk mid-plane density is about $2\times$ smaller than in NE2001, while the peak densities of the Perseus and Car-Sgr arms are increased by factors of $3.7$ and $2$, respectively. These revisions are largely responsible for the improvement in scattering estimation from DM for sightlines through the Galactic plane. The scattering predictions in \model\ are now comparable (if not slightly better) in accuracy to the empirical $\tau$-DM relation used in YMW16 (Figure~\ref{fig:error-hist}).

As with NE2001, there are several important differences between the architecture of \model\ and YMW16. Chief among them is that \model\ forward models scattering from electron density fluctuations, while YMW16 uses the empirical $\tau$-DM relation from \cite{krishnakumar15}. While the $\tau$-DM relation has significant predictive power for Galactic sightlines, it cannot be readily extrapolated to predict the Galactic scattering of extragalactic sources (see below for further discussion). \model\ also includes a number of discrete clumps and voids, while YMW16 only includes clumps and voids for the local ISM and Gum Nebula. 

\begin{figure*}
    \centering
    \includegraphics[width=0.9\textwidth]{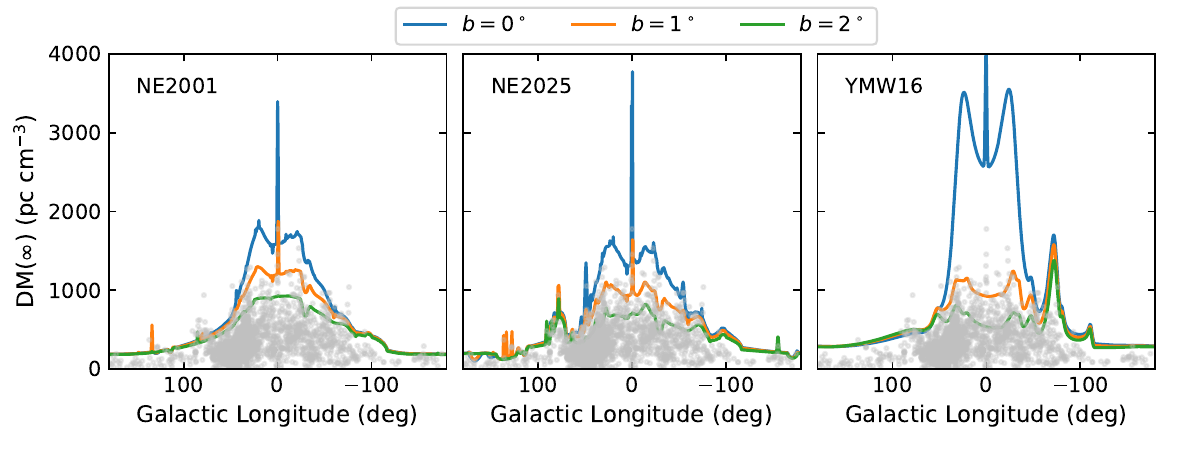}
    \caption{Maximum DM integrated to 30 kpc vs. Galactic longitude predicted by NE2001, \model, and YMW16, at $b=0^\circ$ (blue), $b=1^\circ$ (orange), and $b=2^\circ$ (green). Pulsars at $|b|<2^\circ$ are shown in light grey.}
    \label{fig:all_max_dm}
\end{figure*}

Arguably the most significant difference between \model/NE2001 and YMW16 is the electron density of the thin disk. This difference is readily apparent in Figure~\ref{fig:all_max_dm}, which shows the DM integrated to 30 kpc for NE2001, \model, and YMW16. In the Galactic plane ($b = 0^\circ$), YMW16 predicts about twice as much total electron column due primarily to the larger thin disk density, which is $0.4$ cm$^{-3}$ in YMW16 vs. $0.08$ cm$^{-3}$ in NE2001/\model. The smaller scale height of the thin disk in YMW16 means that this difference in maximum DM is confined to $|b|<1^\circ$. Given degeneracies between the thin disk and spiral arm parameters, it is possible that part of this difference in maximum DM arises from the differences in spiral arm structure adopted by NE2001 and YMW16, as well as differences in the distances used to calibrate the two models' thin disk components. Our own assessment of the thin disk and spiral arms (Section~\ref{sec:spiral}) did not yield evidence supporting a higher thin disk density, but a deeper investigation of these model differences is warranted for future study. Regardless, it is clear that electron density structure within the Galactic plane, and particularly towards the inner Galaxy, is highly challenging to characterize.

\subsection{Consequences for Extragalactic Studies}

\begin{figure*}
    \centering
    \includegraphics[width=\textwidth]{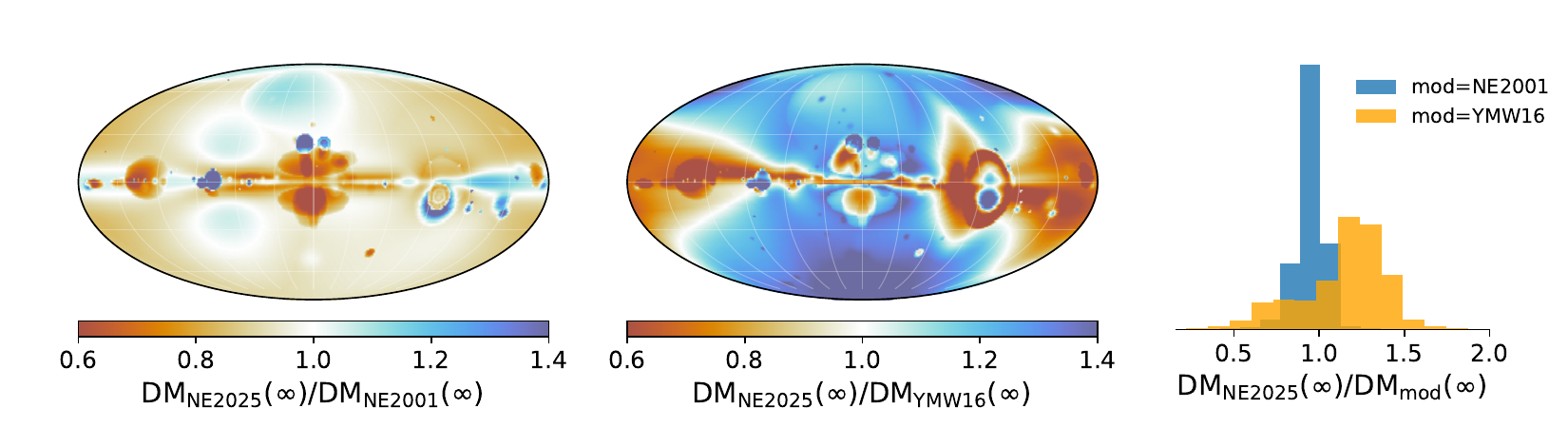}
    \caption{Ratio of the maximum DM integrated through the entire Galaxy in \model\ to the maximum DM in NE2001 (left) and YMW16 (middle), for $10^6$ uniformly distributed sightlines shown as an all-sky map in Galactic coordinates. White grid marks show intervals of $30^\circ$ in longitude and latitude. The far-right panel shows a histogram of the maximum DM ratio, which has a mean and rms of $0.93$ and $0.11$, respectively, for (\model/NE2001) and $1.18$ and $0.26$, respectively, for (\model/YMW16).}
    \label{fig:maxdm_ratios}
\end{figure*}

Extragalactic studies rely on the predicted DM and scattering for sightlines integrated through the entire ISM. Despite our modifications to the large-scale components of NE2001, the total DM integrated through the Galaxy does not change substantially between NE2001 and \model\ for most sightlines. Figure~\ref{fig:maxdm_ratios} shows the ratio of the maximum DM in both models, $\rm DM_{NE2025}(\infty)/DM_{NE2001}(\infty)$, for $10^6$ LOSs distributed uniformly across the sky. On average, $\rm DM_{NE2025}(\infty)/DM_{NE2001}(\infty) = 0.93 \pm 0.11$, although \model\ tends to give even smaller maximum DMs for LOSs through the Fermi Bubbles, and larger maximum DMs along spiral arm tangents. When compared to YMW16, \model\ tends to predict larger maximum DMs; on average, $\rm DM_{NE2025}(\infty)/DM_{YMW16}(\infty) = 1.18 \pm 0.26$, with much of the difference arising from the mid-plane density of the thick disk, leading to YMW16 predicting $n_1H_1 = 18.9$ pc cm$^{-3}$, nearly 10 pc cm$^{-3}$ smaller than the value in \model. 
These changes in maximum DM are especially relevant for FRB studies, but based on initial analysis, we expect that the revisions to previously published FRB DM budgets based on NE2001 will largely be within typically adopted uncertainties on the Galactic DM contribution, with a minority of sightlines being the exception (e.g., FRB 20220319D). 

As in NE2001, \model\ provides scattering predictions for extragalactic sources that explicitly depend on Galactic structure and are properly scaled to account for the differences between plane wave and spherical wave scattering. This approach is preferable to the empirical $\tau$-DM relation based on Galactic pulsars, which tends to overestimate the scattering of extragalactic sources. Extragalactic sources necessarily fall in the ``high-DM'' part of the $\tau$-DM relation, which primarily traces sightlines through the inner Galaxy where the scattering strength per unit DM is larger than in the outer Galaxy and above the Galactic plane (see, e.g., \citealt{ocker2021halo}). We thus expect \model\ to provide more accurate Galactic scattering predictions for extragalactic sightlines, although a detailed assessment is reserved for future work.

\model\ does not include a component for the circumgalactic medium (CGM). At present, uncertainties in the electron density profile of the Galactic CGM lead to a substantial spread in its predicted DM contribution \citep{2019MNRAS.485..648P,2020ApJ...888..105Y,2020MNRAS.496L.106K,das2021,cook2023,ravi2025}. As such, many studies prefer to consider a range of CGM electron density profiles and predicted DMs, which are treated separately from the ISM. Given that the sample of nearby, localized FRBs is expected to grow substantially over the next few years, a future model will be well-positioned to fit for the ISM and CGM electron density structure in tandem, rather than considering these components of Galactic structure in isolation.

\subsection{Directions for a Future Model}

\model\ will be a stepping stone to a completely new Galactic electron density model that considers significant advances in our understanding of ionized gas in and around the Milky Way. There are numerous avenues for improving the modeling framework. Throughout the paper we have noted instances where the assumption of a uniform, statistically homogeneous medium appears to be a poor description of the data (e.g., \revision{in spiral arms}, the Galactic Center, and Cygnus). While we have addressed this issue by explicitly modeling ISM substructure (i.e, clumps and voids) for the most discrepant sightlines, a future model may perform significantly better by directly incorporating gas column densities inferred from multi-wavelength spectral line ratios. \revision{Such efforts will be especially critical for accurate modeling of the spiral arms, the density content of which is likely dominated by \HII regions, many of which are not directly sampled by pulsar sightlines \citep{ocker2024c}. Constraints on the density fluctuations in spiral arms will also be further improved by enlarged scattering measurement samples \citep[e.g.][]{jing2025}.} We will also explore alternative methods to fit the model, including Bayesian frameworks for high-dimensional fitting and uncertainty estimation. 

A future model will explore electron density structures that are not included at present, including an assessment of the CGM and Magellanic Clouds that explicitly considers interactions between these regions through, e.g., the Magellanic Stream and Galactic chimneys. Galactic chimneys (i.e., supernovae-driven outflows) may be a compelling explanation for those pulsars that continue to have DMs overestimated by \model, such as J1509$+$5531 (B1508$+$55), a high-velocity, high-latitude pulsar with a proper motion that traces back to the Cygnus region, in which large numbers of supernovae could have produced a chimney that intersects the pulsar LOS \citep{Chatterjee2005}. This effort will be bolstered not only by upcoming radio surveys that will deliver dramatically expanded samples of pulsars, FRBs, and quasars, but also by all-sky spectroscopic surveys like SDSS-V Local Volume Mapper \citep{drory2024_lvm} and SPHEREx \citep{bock2025}, which are probing Galactic ionized gas in unprecedented detail. Lyman continuum transfer modeling based on stellar parallaxes and 3D dust maps also shows significant promise for inference of 3D ionized gas structure in the local ISM \citep{mccallum2025}. Furthermore, recent advances in Faraday synthesis suggest that simultaneous modeling of electron density and magnetic field strengths, at least for sightlines integrated across the entire Galaxy, is within reach \citep[e.g.][]{hutschenreuter2022}. These developments set the stage for the next generation of Galactic electron density models.

\model\ is provided in both Python and Fortran through the \texttt{mwprop} package, available both on Github\footnote{\url{github.com/stella-ocker/mwprop}} and the Python Package Interface.\footnote{\url{pypi.org/project/mwprop/}} Model usage is described in detail in the software documentation and in Appendix B of \cite{ne20011}.

\acknowledgements{The authors are deeply grateful to Joseph Lazio for his feedback throughout this work. Thanks are also due to the anonymous referee, Vikram Ravi and the DSA-110 science team, David Kaplan, Michael Lam, Chiara Mingarelli, Swarali Patil, Weicong Jing, and Calvin Leung. SKO is supported by the Brinson Foundation through the Brinson Prize Fellowship Program. SKO and JMC are members of the NANOGrav Physics Frontiers Center (NSF award PHY-2020265). This work made use of the \texttt{psrqpy} package \citep{psrqpy} and \texttt{astropy} \citep{astropy2013,astropy2018,astropy2022}. The MeerKAT telescope is operated by the South African Radio Astronomy Observatory, which is a facility of the National Research Foundation, an agency of the Department of Science and Innovation. This publication makes use of data products from the Spectro-Photometer for the History of the Universe, Epoch of Reionization and Ices Explorer (SPHEREx), which is a joint project of the Jet Propulsion Laboratory and the California Institute of Technology, and is funded by the National Aeronautics and Space Administration. Caltech and Carnegie Observatories are located on the traditional and unceded lands of the Tongva people.}


\appendix

\section{Distance Tables}

\startlongtable
\begin{deluxetable*}{L C C C C C c}\label{tab:parallaxes}
\tabletypesize{\scriptsize}
\tablecaption{Pulsar Parallax Distances and Dispersion Measures}
\tablehead{\colhead{Name} & \colhead{$l$ $(^{\circ})$} & \colhead{$b$ $(^{\circ})$} & \colhead{DM (pc cm$^{-3}$)} & \colhead{$\pi$ (mas)} & \colhead{$D$ (kpc)} & \colhead{Ref.}}
\startdata
\rm J0030+0451  & 113.14 & -57.61 & 4.33 &  3.04^{+0.05}_{-0.05}  &  0.33^{+0.01}_{-0.01}  &  \cite{ding2023} \\
\rm J0034-0721  & 110.42 & -69.82 & 10.92 &  0.93^{+0.08}_{-0.07}  &  1.08^{+0.09}_{-0.09}  &  \cite{2009ApJ...698..250C} \\
\rm J0055+5117  & 123.62 & -11.58 & 44.01 &  0.35^{+0.06}_{-0.06}  &  2.86^{+0.59}_{-0.42}  &  \cite{deller2019} \\
\rm J0102+6537  & 124.08 & 2.77 & 65.85 &  0.40^{+0.04}_{-0.04}  &  2.50^{+0.28}_{-0.23}  &  \cite{deller2019} \\
\rm J0108+6608  & 124.65 & 3.33 & 30.55 &  0.47^{+0.04}_{-0.03}  &  2.13^{+0.15}_{-0.17}  &  \cite{deller2019} \\
\rm J0125-2327  & 189.07 & -81.54 & 9.60 &  0.80^{+0.11}_{-0.11}  &  1.25^{+0.20}_{-0.15}  &  \cite{mpta-parallaxes2024} \\
\rm J0139+5814  & 129.22 & -4.04 & 73.81 &  0.37^{+0.04}_{-0.04}  &  2.70^{+0.33}_{-0.26}  &  \cite{2009ApJ...698..250C} \\
\rm J0147+5922  & 130.06 & -2.72 & 40.11 &  0.49^{+0.04}_{-0.09}  &  2.04^{+0.46}_{-0.15}  &  \cite{deller2019} \\
\rm J0157+6212  & 130.59 & 0.33 & 30.21 &  0.55^{+0.04}_{-0.02}  &  1.82^{+0.08}_{-0.12}  &  \cite{deller2019} \\
\rm J0212+5321  & 134.92 & -7.62 & 25.70 &  0.86^{+0.02}_{-0.02}  &  1.16^{+0.03}_{-0.03}  &  \cite{perez2023} \\
\rm J0323+3944  & 152.18 & -14.34 & 26.19 &  1.05^{+0.04}_{-0.04}  &  0.95^{+0.04}_{-0.03}  &  \cite{deller2019} \\
\rm J0332+5434  & 145.00 & -1.22 & 26.76 &  0.61^{+0.01}_{-0.01}  &  1.64^{+0.03}_{-0.03}  &  \cite{kumar2025} \\
\rm J0335+4555  & 150.35 & -8.04 & 47.15 &  0.40^{+0.02}_{-0.03}  &  2.50^{+0.18}_{-0.13}  &  \cite{deller2019} \\
\rm J0357+5236  & 149.10 & -0.52 & 103.71 &  0.30^{+0.03}_{-0.08}  &  3.33^{+1.15}_{-0.29}  &  \cite{deller2019} \\
\rm J0358+5413  & 148.19 & 0.81 & 57.14 &  0.91^{+0.16}_{-0.16}  &  1.10^{+0.23}_{-0.16}  &  \cite{chatterjee2004} \\
\rm J0437-4715  & 253.39 & -41.96 & 2.64 &  6.43^{+0.04}_{-0.04}  &  0.155^{+0.001}_{-0.001}  &  \cite{reardon2024} \\
\rm J0454+5543  & 152.62 & 7.55 & 14.59 &  0.84^{+0.04}_{-0.05}  &  1.19^{+0.08}_{-0.05}  &  \cite{2009ApJ...698..250C} \\
\rm J0509+3801  & 168.27 & -1.19 & 69.08 &  0.27^{+0.02}_{-0.02}  &  3.70^{+0.30}_{-0.26}  &  \cite{ding2024b} \\
\rm J0534+2200  & 184.56 & -5.78 & 56.77 &  0.53^{+0.06}_{-0.06}  &  1.89^{+0.24}_{-0.19}  &  \cite{nanograv2023} \\
\rm J0538+2817  & 179.72 & -1.69 & 39.57 &  0.72^{+0.12}_{-0.09}  &  1.39^{+0.20}_{-0.20}  &  \cite{2009ApJ...698..250C} \\
\rm J0601-0527  & 212.20 & -13.48 & 80.54 &  0.48^{+0.04}_{-0.04}  &  2.08^{+0.19}_{-0.16}  &  \cite{deller2019} \\
\rm J0610-2100  & 227.75 & -18.18 & 60.67 &  0.73^{+0.10}_{-0.10}  &  1.37^{+0.22}_{-0.17}  &  \cite{ding2023} \\
\rm J0613-0200  & 210.41 & -9.30 & 38.77 &  1.00^{+0.05}_{-0.05}  &  1.00^{+0.05}_{-0.05}  &  \cite{epta2023}\\
\rm J0614+2229  & 188.79 & 2.39 & 96.91 &  0.28^{+0.02}_{-0.03}  &  3.57^{+0.44}_{-0.26}  &  \cite{deller2019} \\
\rm J0621+1002  & 200.57 & -2.01 & 36.55 &  0.74^{+0.14}_{-0.14}  &  1.35^{+0.32}_{-0.21}  &  \cite{ding2023} \\
\rm J0629+2415  & 188.82 & 6.22 & 84.18 &  0.33^{+0.04}_{-0.05}  &  3.03^{+0.54}_{-0.33}  &  \cite{deller2019} \\
\rm J0630-2834  & 236.95 & -16.76 & 34.42 &  3.00^{+0.40}_{-0.40}  &  0.33^{+0.05}_{-0.04}  &  \cite{deller2009} \\
\rm J0636+5128  & 163.91 & 18.64 & 11.11 &  0.70^{+0.17}_{-0.17}  &  1.43^{+0.46}_{-0.28}  &  \cite{nanograv2023} \\
\rm J0636-3044  & 239.68 & -16.34 & 15.46 &  4.30^{+0.70}_{-0.70}  &  0.23^{+0.05}_{-0.03}  &  \cite{mpta-parallaxes2024} \\
\rm J0645+5158  & 163.96 & 20.25 & 18.25 &  0.60^{+0.10}_{-0.10}  &  1.67^{+0.33}_{-0.24}  &  \cite{nanograv2023} \\
\rm J0659+1414  & 201.11 & 8.26 & 13.94 &  3.47^{+0.36}_{-0.36}  &  0.29^{+0.03}_{-0.03}  &  \cite{brisken2003} \\
\rm J0729-1836  & 233.76 & -0.34 & 61.29 &  0.49^{+0.10}_{-0.08}  &  2.04^{+0.40}_{-0.35}  &  \cite{deller2019} \\
\rm J0737-3039A  & 245.24 & -4.50 & 48.92 &  1.30^{+0.13}_{-0.11}  &  0.77^{+0.07}_{-0.07}  &  \cite{kramer2021} \\
\rm J0740+6620  & 149.73 & 29.60 & 14.96 &  1.12^{+0.22}_{-0.22}  &  0.89^{+0.22}_{-0.15}  &  \cite{nanograv2023} \\
\rm J0751+1807  & 202.73 & 21.09 & 30.24 &  0.80^{+0.04}_{-0.04}  &  1.25^{+0.07}_{-0.06}  &  \cite{epta2023}\\
\rm J0814+7429  & 140.00 & 31.62 & 5.75 &  2.31^{+0.04}_{-0.04}  &  0.43^{+0.01}_{-0.01}  &  \cite{brisken2002} \\
\rm J0820-1350  & 235.89 & 12.59 & 40.94 &  0.51^{+0.03}_{-0.04}  &  1.96^{+0.17}_{-0.11}  &  \cite{2009ApJ...698..250C} \\
\rm J0826+2637  & 196.96 & 31.74 & 19.48 &  2.01^{+0.01}_{-0.01}  &  0.497^{+0.002}_{-0.003}  &  \cite{deller2019} \\
\rm J0835-4510  & 263.55 & -2.79 & 67.77 &  3.50^{+0.20}_{-0.20}  &  0.29^{+0.02}_{-0.02}  &  \cite{dodson2003} \\
\rm J0837+0610  & 219.72 & 26.27 & 12.86 &  1.63^{+0.15}_{-0.15}  &  0.61^{+0.06}_{-0.05}  &  \cite{liu2016} \\
\rm J0922+0638  & 225.42 & 36.39 & 27.30 &  0.83^{+0.13}_{-0.13}  &  1.20^{+0.22}_{-0.16}  &  \cite{2001ApJ...550..287C} \\
\rm J0953+0755  & 228.91 & 43.70 & 2.97 &  3.82^{+0.07}_{-0.07}  &  0.262^{+0.005}_{-0.005}  &  \cite{brisken2002} \\
\rm J1012+5307  & 160.35 & 50.86 & 9.02 &  1.14^{+0.04}_{-0.04}  &  0.88^{+0.03}_{-0.03}  &  \cite{ding2023} \\
\rm J1022+1001  & 231.79 & 51.10 & 10.26 &  1.38^{+0.04}_{-0.03}  &  0.72^{+0.02}_{-0.02}  &  \cite{deller2019} \\
\rm J1023+0038  & 243.49 & 45.78 & 14.32 &  0.73^{+0.02}_{-0.02}  &  1.37^{+0.04}_{-0.04}  &  \cite{2012ApJ...756L..25D}\\
\rm J1024-0719  & 251.70 & 40.52 & 6.49 &  0.93^{+0.05}_{-0.05}  &  1.08^{+0.06}_{-0.05}  &  \cite{ding2023} \\
\rm J1136+1551  & 241.90 & 69.20 & 4.84 &  2.705^{+0.009}_{-0.009}  &  0.370^{+0.001}_{-0.001}  &  \cite{kumar2025} \\
\rm J1239+2453  & 252.45 & 86.54 & 9.25 &  1.16^{+0.08}_{-0.08}  &  0.86^{+0.06}_{-0.06}  &  \cite{brisken2002} \\
\rm J1302-6350  & 304.18 & -0.99 & 146.73 &  0.41^{+0.03}_{-0.03}  &  2.44^{+0.19}_{-0.17}  &  \cite{2018ApJ...864...26J} \\
\rm J1321+8323  & 121.89 & 33.67 & 13.32 &  0.97^{+0.04}_{-0.14}  &  1.03^{+0.17}_{-0.04}  &  \cite{deller2019} \\
\rm J1455-3330  & 330.72 & 22.56 & 13.57 &  1.30^{+0.10}_{-0.10}  &  0.77^{+0.06}_{-0.05}  &  \cite{epta2023}\\
\rm J1456-6843  & 313.87 & -8.54 & 8.61 &  2.20^{+0.30}_{-0.30}  &  0.45^{+0.07}_{-0.05}  &  \cite{bailes90} \\
\rm J1509+5531  & 91.33 & 52.29 & 19.62 &  0.47^{+0.03}_{-0.03}  &  2.13^{+0.15}_{-0.13}  &  \cite{2009ApJ...698..250C} \\
\rm J1518+4904  & 80.81 & 54.28 & 11.61 &  1.23^{+0.04}_{-0.03}  &  0.81^{+0.02}_{-0.02}  &  \cite{ding2023} \\
\rm J1532+2745  & 43.48 & 54.50 & 14.69 &  0.62^{+0.03}_{-0.10}  &  1.61^{+0.31}_{-0.07}  &  \cite{deller2019} \\
\rm J1537+1155  & 19.85 & 48.34 & 11.62 &  0.96^{+0.01}_{-0.01}  &  1.04^{+0.01}_{-0.01}  &  \cite{fonseca2014} \\
\rm J1543+0929  & 17.81 & 45.78 & 34.98 &  0.13^{+0.02}_{-0.02}  &  7.69^{+1.40}_{-1.03}  &  \cite{2009ApJ...698..250C} \\
\rm J1543-0620  & 0.57 & 36.61 & 18.30 &  0.32^{+0.03}_{-0.04}  &  3.12^{+0.51}_{-0.25}  &  \cite{deller2019} \\
\rm J1559-4438  & 334.54 & 6.37 & 55.94 &  0.38^{+0.08}_{-0.08}  &  2.63^{+0.71}_{-0.46}  &  \cite{deller2009} \\
\rm J1600-3053  & 344.09 & 16.45 & 52.33 &  0.70^{+0.02}_{-0.02}  &  1.43^{+0.04}_{-0.04}  &  \cite{epta2023}\\
\rm J1607-0032  & 10.72 & 35.47 & 10.68 &  0.91^{+0.03}_{-0.05}  &  1.10^{+0.06}_{-0.03}  &  \cite{deller2019} \\
\rm J1614-2230  & 352.64 & 20.19 & 34.48 &  1.30^{+0.09}_{-0.09}  &  0.77^{+0.06}_{-0.05}  &  \cite{2016AA...587A.109G} \\
\rm J1623-0908  & 5.30 & 27.18 & 68.18 &  0.59^{+0.10}_{-0.10}  &  1.69^{+0.35}_{-0.25}  &  \cite{deller2019} \\
\rm J1629-6902  & 320.37 & -13.93 & 29.49 &  0.90^{+0.20}_{-0.20}  &  1.11^{+0.32}_{-0.20}  &  \cite{mpta-parallaxes2024} \\
\rm J1640+2224  & 41.05 & 38.27 & 18.43 &  0.73^{+0.06}_{-0.06}  &  1.37^{+0.12}_{-0.10}  &  \cite{ding2023} \\
\rm J1643-1224  & 5.67 & 21.22 & 62.40 &  1.10^{+0.10}_{-0.10}  &  0.91^{+0.09}_{-0.08}  &  \cite{ding2023} \\
\rm J1645-0317  & 14.11 & 26.06 & 35.76 &  0.25^{+0.03}_{-0.02}  &  4.00^{+0.33}_{-0.40}  &  \cite{deller2019} \\
\rm J1658-5324  & 334.87 & -6.63 & 30.83 &  1.30^{+0.30}_{-0.30}  &  0.77^{+0.23}_{-0.14}  &  \cite{mpta-parallaxes2024} \\
\rm J1703-1846  & 3.23 & 13.56 & 49.55 &  0.35^{+0.05}_{-0.05}  &  2.86^{+0.48}_{-0.36}  &  \cite{deller2019} \\
\rm J1713+0747  & 28.75 & 25.22 & 15.99 &  0.80^{+0.01}_{-0.01}  &  1.25^{+0.02}_{-0.02}  &  \cite{epta2023} \\
\rm J1723-2837  & 357.62 & 4.26 & 19.69 &  1.07^{+0.05}_{-0.05}  &  0.93^{+0.05}_{-0.04}  &  \cite{2018ApJ...864...26J} \\
\rm J1730-2304  & 3.14 & 6.02 & 9.63 &  2.00^{+0.06}_{-0.06}  &  0.50^{+0.02}_{-0.01}  &  \cite{epta2023} \\
\rm J1738+0333  & 27.72 & 17.74 & 33.77 &  0.50^{+0.06}_{-0.06}  &  2.00^{+0.27}_{-0.21}  &  \cite{ding2023} \\
\rm J1741-0840  & 16.96 & 11.30 & 74.90 &  0.28^{+0.05}_{-0.06}  &  3.57^{+0.97}_{-0.54}  &  \cite{deller2019} \\
\rm J1744-1134  & 14.79 & 9.18 & 3.14 &  2.50^{+0.03}_{-0.03}  &  0.400^{+0.005}_{-0.005}  &  \cite{epta2023} \\
\rm J1754+5201  & 79.61 & 29.63 & 35.01 &  0.16^{+0.03}_{-0.02}  &  6.25^{+1.00}_{-0.96}  &  \cite{deller2019} \\
\rm J1809-1943  & 10.73 & -0.16 & 178.00 &  0.40^{+0.05}_{-0.05}  &  2.50^{+0.36}_{-0.28}  &  \cite{ding2020} \\
\rm J1811-2405  & 7.07 & -2.56 & 60.62 &  0.70^{+0.15}_{-0.15}  &  1.43^{+0.39}_{-0.25}  &  \cite{mpta-parallaxes2024} \\
\rm J1818-1607  & 14.81 & -0.14 & 699.00 &  0.12^{+0.02}_{-0.02}  &  8.33^{+1.67}_{-1.19}  &  \cite{ding2024} \\
\rm J1820-0427  & 25.46 & 4.73 & 84.44 &  0.35^{+0.05}_{-0.06}  &  2.86^{+0.59}_{-0.36}  &  \cite{deller2019} \\
\rm J1832-0836  & 23.11 & 0.26 & 28.19 &  0.50^{+0.11}_{-0.11}  &  2.00^{+0.56}_{-0.36}  &  \cite{nanograv2023} \\
\rm J1833-0338  & 27.66 & 2.27 & 234.54 &  0.41^{+0.05}_{-0.07}  &  2.44^{+0.50}_{-0.27}  &  \cite{deller2019} \\
\rm J1840+5640  & 86.08 & 23.82 & 26.77 &  0.65^{+0.07}_{-0.01}  &  1.54^{+0.02}_{-0.14}  &  \cite{deller2019} \\
\rm J1853+1303  & 44.87 & 5.37 & 30.57 &  0.53^{+0.07}_{-0.07}  &  1.89^{+0.29}_{-0.22}  &  \cite{ding2023} \\
\rm J1857+0943  & 42.29 & 3.06 & 13.30 &  0.80^{+0.06}_{-0.06}  &  1.25^{+0.10}_{-0.09}  &  \cite{epta2023} \\
\rm J1901-0906  & 25.98 & -6.44 & 72.68 &  0.51^{+0.07}_{-0.04}  &  1.96^{+0.17}_{-0.24}  &  \cite{deller2019} \\
\rm J1909-3744  & 359.73 & -19.60 & 10.39 &  0.80^{+0.01}_{-0.01}  &  1.25^{+0.02}_{-0.02}  &  \cite{ppta2021} \\
\rm J1910+1256  & 46.56 & 1.80 & 38.07 &  0.25^{+0.04}_{-0.04}  &  4.00^{+0.65}_{-0.49}  &  \cite{ding2023} \\
\rm J1911+1347  & 47.52 & 1.81 & 30.99 &  0.40^{+0.09}_{-0.09}  &  2.50^{+0.73}_{-0.46}  &  \cite{epta2023} \\
\rm J1913+1400  & 47.88 & 1.59 & 145.05 &  0.18^{+0.03}_{-0.02}  &  5.56^{+0.81}_{-0.72}  &  \cite{deller2019} \\
\rm J1918-0642  & 30.03 & -9.12 & 26.59 &  0.71^{+0.07}_{-0.07}  &  1.41^{+0.15}_{-0.13}  &  \cite{ding2023} \\
\rm J1923+2515  & 58.95 & 4.75 & 18.86 &  1.00^{+0.24}_{-0.24}  &  1.00^{+0.32}_{-0.19}  &  \cite{nanograv2023} \\
\rm J1932+1059  & 47.38 & -3.88 & 3.18 &  2.77^{+0.07}_{-0.07}  &  0.36^{+0.01}_{-0.01}  &  \cite{chatterjee2004} \\
\rm J1937+2544  & 60.84 & 2.27 & 53.22 &  0.31^{+0.03}_{-0.03}  &  3.23^{+0.33}_{-0.29}  &  \cite{deller2019} \\
\rm J1939+2134  & 57.51 & -0.29 & 71.02 &  0.35^{+0.03}_{-0.03}  &  2.86^{+0.27}_{-0.23}  &  \cite{ding2023} \\
\rm J1946-5403  & 343.88 & -29.58 & 23.73 &  0.80^{+0.10}_{-0.11}  &  1.25^{+0.20}_{-0.14}  &  \cite{mpta-parallaxes2024} \\
\rm J1959+2048  & 59.20 & -4.70 & 29.11 &  0.39^{+0.09}_{-0.10}  &  2.56^{+0.88}_{-0.48}  &  \cite{romani2022} \\
\rm J2006-0807  & 34.10 & -20.30 & 32.39 &  0.42^{+0.01}_{-0.10}  &  2.38^{+0.75}_{-0.06}  &  \cite{deller2019} \\
\rm J2010-1323  & 29.45 & -23.54 & 22.16 &  0.50^{+0.11}_{-0.11}  &  2.00^{+0.56}_{-0.36}  &  \cite{nanograv2023} \\
\rm J2018+2839  & 68.10 & -3.98 & 14.20 &  1.03^{+0.10}_{-0.10}  &  0.97^{+0.10}_{-0.09}  &  \cite{brisken2002} \\
\rm J2022+5154  & 87.86 & 8.38 & 22.55 &  0.50^{+0.07}_{-0.07}  &  2.00^{+0.33}_{-0.25}  &  \cite{brisken2002} \\
\rm J2032+4127  & 80.22 & 1.03 & 114.67 &  0.69^{+0.03}_{-0.03}  &  1.45^{+0.07}_{-0.07}  &  \cite{2018ApJ...864...26J} \\
\rm J2043+1711  & 61.92 & -15.31 & 20.71 &  0.60^{+0.04}_{-0.04}  &  1.67^{+0.13}_{-0.11}  &  \cite{nanograv2023} \\
\rm J2046-0421  & 42.68 & -27.39 & 35.80 &  0.16^{+0.03}_{-0.04}  &  6.25^{+2.22}_{-0.87}  &  \cite{deller2019} \\
\rm J2048-1616  & 30.51 & -33.08 & 11.46 &  1.05^{+0.03}_{-0.02}  &  0.95^{+0.02}_{-0.03}  &  \cite{2009ApJ...698..250C} \\
\rm J2055+3630  & 79.13 & -5.59 & 97.42 &  0.17^{+0.03}_{-0.03}  &  5.88^{+1.26}_{-0.88}  &  \cite{2009ApJ...698..250C} \\
\rm J2113+2754  & 74.99 & -14.03 & 25.11 &  0.70^{+0.02}_{-0.02}  &  1.43^{+0.05}_{-0.05}  &  \cite{deller2019} \\
\rm J2113+4644  & 89.00 & -1.27 & 141.26 &  0.45^{+0.08}_{-0.07}  &  2.22^{+0.41}_{-0.34}  &  \cite{deller2019} \\
\rm J2124-3358  & 10.92 & -45.44 & 4.60 &  2.10^{+0.10}_{-0.10}  &  0.48^{+0.02}_{-0.02}  &  \cite{epta2023} \\
\rm J2129-0429  & 48.91 & -36.94 & 16.88 &  0.42^{+0.09}_{-0.09}  &  2.38^{+0.63}_{-0.41}  &  \cite{2018ApJ...864...26J} \\
\rm J2144-3933  & 2.79 & -49.47 & 3.35 &  6.05^{+0.56}_{-0.56}  &  0.17^{+0.02}_{-0.01}  &  \cite{deller2009} \\
\rm J2145-0750  & 47.78 & -42.08 & 9.00 &  1.60^{+0.06}_{-0.01}  &  0.625^{+0.004}_{-0.02}  &  \cite{deller2019} \\
\rm J2149+6329  & 104.25 & 7.41 & 129.72 &  0.36^{+0.07}_{-0.06}  &  2.78^{+0.56}_{-0.45}  &  \cite{deller2019} \\
\rm J2157+4017  & 90.49 & -11.34 & 71.12 &  0.28^{+0.06}_{-0.06}  &  3.57^{+0.97}_{-0.63}  &  \cite{2009ApJ...698..250C} \\
\rm J2222-0137  & 62.02 & -46.08 & 3.28 &  3.72^{+0.01}_{-0.01}  &  0.269^{+0.001}_{-0.001}  &  \cite{ding2024} \\
\rm J2225+6535  & 108.64 & 6.85 & 36.44 &  1.20^{+0.17}_{-0.20}  &  0.83^{+0.17}_{-0.10}  &  \cite{deller2019} \\
\rm J2234+0611  & 72.99 & -43.01 & 10.77 &  0.80^{+0.10}_{-0.10}  &  1.25^{+0.18}_{-0.14}  &  \cite{nanograv2023} \\
\rm J2241-5236  & 337.46 & -54.93 & 11.41 &  0.90^{+0.04}_{-0.04}  &  1.11^{+0.05}_{-0.05}  &  \cite{ppta2021} \\
\rm J2248-0101  & 69.26 & -50.62 & 29.05 &  0.26^{+0.05}_{-0.07}  &  3.85^{+1.42}_{-0.62}  &  \cite{deller2019} \\
\rm J2305+3100  & 97.72 & -26.66 & 49.58 &  0.22^{+0.03}_{-0.03}  &  4.55^{+0.66}_{-0.59}  &  \cite{deller2019} \\
\rm J2313+4253  & 104.41 & -16.42 & 17.28 &  0.93^{+0.06}_{-0.07}  &  1.08^{+0.09}_{-0.07}  &  \cite{2009ApJ...698..250C} \\
\rm J2317+1439  & 91.36 & -42.36 & 21.90 &  0.60^{+0.12}_{-0.12}  &  1.67^{+0.42}_{-0.28}  &  \cite{nanograv2023} \\
\rm J2317+2149  & 95.83 & -36.07 & 20.87 &  0.51^{+0.06}_{-0.05}  &  1.96^{+0.21}_{-0.21}  &  \cite{deller2019} \\
\rm J2322+2057  & 96.51 & -37.31 & 13.38 &  1.00^{+0.21}_{-0.21}  &  1.00^{+0.27}_{-0.17}  &  \cite{nanograv2023} \\
\rm J2346-0609  & 83.80 & -64.01 & 22.50 &  0.27^{+0.02}_{-0.04}  &  3.70^{+0.57}_{-0.27}  &  \cite{deller2019} \\
\rm J2354+6155  & 116.24 & -0.19 & 94.66 &  0.41^{+0.03}_{-0.04}  &  2.44^{+0.26}_{-0.17}  &  \cite{deller2019} \\
\enddata
\tablecomments{Left to right: Pulsar name, Galactic longitude and latitude, DM, parallax, distance, and reference.} 
\end{deluxetable*}

\startlongtable
\begin{deluxetable*}{l C C C C C}\label{tab:globs}
\tabletypesize{\scriptsize}
\tablecaption{Globular Cluster Distances and Dispersion Measures}
\tablehead{\colhead{Name} & \colhead{$l$ $(^{\circ})$} & \colhead{$b$ $(^{\circ})$} & \colhead{DM (pc cm$^{-3}$)} & \colhead{$D$ (kpc)} & \colhead{No. Pulsars}}
\startdata
NGC 6522 & 1.02 & -3.90 & 193.26\pm0.58 & 7.29\pm0.21 & 6 \\
NGC 6652 & 1.53 & -11.38 & 63.43\pm0.08 & 9.46\pm0.14 & 2 \\
NGC 6637 & 1.72 & -10.27 & 82.00\pm0.10 & 8.90\pm0.10 & 2 \\
NGC 6624 & 2.79 & -7.91 & 88.10\pm0.87 & 8.02\pm0.11 & 12 \\
NGC 6681 & 2.81 & -12.51 & 70.93\pm0.17 & 9.36\pm0.11 & 3 \\
Ter 5 & 3.84 & 1.69 & 237.93\pm0.26 & 6.62\pm0.15 & 49 \\
NGC 5904 & 3.86 & 46.80 & 29.47\pm0.10 & 7.48\pm0.06 & 7 \\
NGC 6342 & 4.90 & 9.72 & 71.20\pm0.20 & 8.01\pm0.23 & 2 \\
NGC 6544 & 5.84 & -2.20 & 135.88\pm0.96 & 2.58\pm0.06 & 3 \\
NGC 6440 & 7.73 & 3.80 & 222.37\pm0.97 & 8.25\pm0.24 & 8 \\
NGC 6626 & 7.80 & -5.58 & 119.99\pm0.33 & 5.37\pm0.10 & 14 \\
NGC 6656 & 9.89 & -7.55 & 89.95\pm1.17 & 3.30\pm0.04 & 4 \\
NGC 6254 & 15.14 & 23.08 & 43.63\pm0.27 & 5.07\pm0.06 & 2 \\
NGC 6218 & 15.72 & 26.31 & 42.60\pm0.10 & 5.11\pm0.05 & 2 \\
NGC 6517 & 19.23 & 6.76 & 182.35\pm0.52 & 9.23\pm0.56 & 21 \\
NGC 6539 & 20.80 & 6.78 & 186.38\pm0.01 & 8.16\pm0.39 & 1 \\
NGC 6402 & 21.32 & 14.81 & 80.46\pm0.55 & 9.14\pm0.25 & 5 \\
NGC 6712 & 25.35 & -4.32 & 155.12 & 7.38\pm0.24 & 1 \\
NGC 7099 & 27.18 & -46.84 & 25.0635\pm0.0005 & 8.46\pm0.09 & 2 \\
Glimpse C01 & 31.31 & -0.10 & 486.72\pm8.23 & 3.40\pm0.10 & 6 \\
NGC 6760 & 36.11 & -3.92 & 199.68\pm2.99 & 8.41\pm0.43 & 2 \\
NGC 6749 & 36.20 & -2.21 & 193.69 & 7.59\pm0.21 & 1 \\
NGC 5272 & 42.22 & 78.71 & 26.38\pm0.06 & 10.18\pm0.08 & 5 \\
NGC 7089 & 53.37 & -35.77 & 43.62\pm0.10 & 11.69\pm0.11 & 10 \\
NGC 6838 & 56.75 & -4.56 & 166.95\pm1.10 & 4.00\pm0.05 & 5 \\
NGC 6205 & 59.01 & 40.91 & 30.26\pm0.17 & 7.42\pm0.08 & 9 \\
NGC 7078 & 65.01 & -27.31 & 66.91\pm0.16 & 10.71\pm0.10 & 15 \\
NGC 6341 & 68.34 & 34.86 & 35.40\pm0.05 & 8.50\pm0.07 & 2 \\
NGC 1904 & 227.23 & -29.35 & 62.30 & 13.08\pm0.18 & 1 \\
NGC 1851 & 244.51 & -35.03 & 51.85\pm0.12 & 11.95\pm0.13 & 15 \\
NGC 2808 & 282.19 & -11.25 & 108.96\pm0.20 & 10.06\pm0.11 & 4 \\
NGC 362 & 301.53 & -46.25 & 24.80\pm0.10 & 8.83\pm0.10 & 12 \\
NGC 104 & 305.90 & -44.89 & 24.36\pm0.04 & 4.52\pm0.03 & 42 \\
NGC 5139 & 309.10 & 14.97 & 98.70\pm0.64 & 5.43\pm0.05 & 19 \\
NGC 5024 & 332.96 & 79.76 & 25.51\pm0.31 & 18.50\pm0.18 & 5 \\
NGC 6752 & 336.49 & -25.63 & 33.32\pm0.05 & 4.12\pm0.04 & 9 \\
NGC 5986 & 337.02 & 13.27 & 92.17 & 10.54\pm0.13 & 1 \\
NGC 6397 & 338.17 & -11.96 & 72.00\pm0.20 & 2.48\pm0.02 & 2 \\
NGC 6121 & 350.97 & 15.97 & 62.86 & 1.85\pm0.02 & 1 \\
NGC 6093 & 352.67 & 19.45 & 66.71\pm0.10 & 10.34\pm0.12 & 3 \\
NGC 6441 & 353.53 & -5.01 & 231.05\pm0.76 & 12.73\pm0.16 & 9 \\
NGC 6266 & 353.57 & 7.32 & 113.80\pm0.28 & 6.03\pm0.09 & 10 \\
NGC 6316 & 357.18 & 5.76 & 172.10 & 11.15\pm0.39 & 1 \\
Ter 1 & 357.57 & 1.00 & 381.00\pm0.52 & 5.67\pm0.17 & 8 \\
Ter 6 & 358.57 & -2.16 & 383.08 & 7.27\pm0.35 & 1 \\
\enddata
\tablecomments{Left to right: Globular cluster name, Galactic longitude, Galactic latitude, DM, distance, and number of pulsars in the cluster. Errors on DM represent the standard error on the mean of all pulsar DMs in that cluster -- for clusters with a single pulsar, no error is quoted. Distances are from \cite{baumgardt2021}.} 
\end{deluxetable*}

\section{Fitting Functions}\label{app:fitting-functions}

Here we elaborate on the functional forms for the fitting criteria described in Section~\ref{sec:fitting}. To fit for parameters $\Theta$ by minimizing the difference between observed and predicted distances and DMs, we use a logarithmic likelihood function of the form
\begin{multline}
    \mathcal{L}(\Theta) = \prod_j \int f_{D_j}(D_j) dD_j \\ \times {\rm exp}[-({\rm ln\ DM_j} - {\rm ln\ }N_{e_j})^2/2\sigma_{{\rm ln}N_{e_j}}^2] 
\end{multline}
where $D_j$ and $\rm DM_j$ are the observed distance and DM, respectively, for the $j$-th pulsar and $N_e$ is the modeled DM. The likelihood function marginalizes over the PDF for distance, which we assume to be Gaussian for globular cluster distances. For parallax distances, the PDF is non-Gaussian and takes the form 
\begin{equation}
    f_{D_j}(D_j) = \frac{1}{D_j^2 \sqrt{2\pi \sigma_{\pi_j}^2}}{\rm exp}[-((1/D) - (1/D_j))^2/2\sigma_{\pi_j}^2]
\end{equation}
where $\pi_j$ is the measured parallax \citep[e.g.][]{ocker2020}. 

Where we use $\chi^2$ minimization, we adopt the definition 
\begin{equation}
    \chi^2 = \sum_j (\hat{\theta} - \theta_{j})^2/{\theta_j}
\end{equation}
where $\hat{\theta}$ is the modeled estimate of a given observable $\theta$.

For scattering measurements, we specifically use
\begin{equation}
    \chi_\tau^2 = \frac{1}{n} \sum_{j=1}^n \Delta_\tau^2
\end{equation}
where $\Delta_\tau = {\rm log}_{10}(\tau_{\rm obs}/\tau_{\rm mod})$ and $\tau$ is the pulse broadening delay. For pulsars with only measured scintillation bandwidths $\Delta \nu_{\rm d}$, we first convert $\Delta \nu_{\rm d}$ to $\tau_{\rm obs}$ using $1.16 = 2\pi\tau\Delta \nu_{\rm d}$ \citep{cordesrickett98,ne20011}.

The final fitting criterion used is based on minimizing the number of pulsars with DMs greater than the model maximum, $\Nxgal$, which is accomplished by finding the model parameter values $\Theta$ for which the following function is maximized: 
\begin{equation}
    f(\Theta) = \frac{d^2\Nxgal}{d\Nxgal^2}.
\end{equation}

\newpage

\section{Clumps and Voids Tables}

\startlongtable
\begin{deluxetable*}{C C C C C C C l L}\label{tab:clumps}
\tabletypesize{\scriptsize}
\tablecaption{Clumps New or Revised in \model}
\tablehead{\colhead{$l$ $(^{\circ})$} & \colhead{$b$ $(^{\circ})$} & \colhead{$n_{e}$ (cm$^{-3}$)} & \colhead{$F_c$} & \colhead{$D_c$ (kpc)} & \colhead{$r_c$ (kpc)} &  \colhead{Edge} & \colhead{Type} & \colhead{Name}}
\startdata
-179.057 & 65.296 & 0.609 & 0.0 & 3.302 & 0.01 & 0 & PX & \rm J1105+37 \\
-175.120 & -1.730 & 1.6 & 0.0 & 0.48 & 0.02 & 0 & PX & \rm J0557+2442\_P(HII) \\
-164.824 & -40.525 & 1.14 & 0.0 & 4.617 & 0.01 & 0 & PX & \rm J0357-05 \\
-158.892 & 8.258 & 0.497 & 0.06 & 0.144 & 0.01 & 0 & P & \rm J0659+1414 \\
-155.729 & 2.200 & 5.0 & 13.0 & 1.508 & 0.025 & 0 & MPX & \rm J0646+0905 \\
-150.960 & -19.500 & 1.7 & 0.0 & 0.437 & 0.02 & 0 & PX & \rm J0529-0715(HII) \\
-149.587 & -9.305 & 0.40 & 0.0 & 0.288 & 0.01 & 0 & P & \rm J0613-0200 \\
-148.925 & -32.629 & 0.1 & 57.0 & 1.634 & 0.01 & 0 & P & \rm J0450-1248 \\
-146.296 & -12.606 & 0.493 & 0.19 & 0.61 & 0.05 & 0 & P & \rm J0601-0527(HII) \\
-143.773 & -22.957 & 1.1 & 0.0 & 7.692 & 0.01 & 0 & PX & \rm J0534-13 \\
-138.742 & 50.980 & 0.731 & 0.0 & 3.861 & 0.01 & 0 & PX & \rm J1010+15 \\
-132.253 & -18.184 & 1.964 & 0.0 & 0.685 & 0.01 & 0 & P & \rm J0610-2100 \\
-122.999 & -12.645 & 2.688 & 0.0 & 0.500 & 0.01 & 1 & PX & \rm J0648-27 \\
-121.683 & -16.551 & 4.257 & 0.00075 & 0.167 & 0.006 & 0 & MP & \rm J0630-2834 \\
-121.066 & -5.826 & 6.091 & 0.0 & 0.500 & 0.01 & 1 & PX & \rm J0719-2545 \\
-112.547 & 9.799 & 0.80 & 0.0 & 0.500 & 0.01 & 1 & PX & \rm J0837-24 \\
-106.068 & 20.138 & 0.855 & 0.0 & 0.5 & 0.01 & 1 & PX & \rm J0930-2301 \\
-100.000 & -1.000 & 0.430 & 0.71 & 0.500 & 0.145 & 1 & P & \rm GumI \\
-97.200 & 2.700 & 0.500 & 1.20 & 0.500 & 0.032 & 1 & P & \rm GumII \\
-96.750 & -9.000 & 4.000 & 1.90 & 0.300 & 0.044 & 0 & P & \rm VelaIras \\
-99.810 & -18.723 & 1.474 & 0.0 & 0.500 & 0.01 & 1 & PX & \rm J0702-4956 \\
-92.884 & -1.193 & 5.906 & 0.0 & 1.24 & 0.01 & 0 & PX & \rm J0855-4658(HII) \\
-91.298 & 12.033 & 0.62 & 0.0 & 0.5 & 0.01 & 1 & PX & \rm J0952-3839 \\
-74.564 & 0.008 & 12.859 & 0.0 & 3.75 & 0.007 & 0 & PX & \rm J1032-5804(HII) \\
-60.957 & 0.186 & 4.8 & 0.0 & 9.88 & 0.028 & 0 & MPX & \rm G299.043+0.186(HII) \\
-55.540 & -3.460 & 0.55 & 3.9 & 6.580 & 0.09 & 1 & PX & \rm J1306-6617 \\
-54.600 & -1.000 & 2.1 & 1.50 & 4.7 & 0.123 & 0 & MPX & \rm G305(HII) \\
-54.152 & 0.191 & 2.1 & 50.00 & 4.7 & 0.01 & 0 & P & \rm J1316-6232(HII) \\
-48.150 & -0.540 & 2.2 & 0.0 & 1.5 & 0.02 & 0 & MPX & \rm RCW83(HII) \\
-39.890 & -0.510 & 0.72 & 130.0 & 0.93 & 0.027 & 0 & MP & \rm J1522-5829(HII) \\
-39.326 & -25.650 & 2.042 & 0.0 & 6.93 & 0.01 & 0 & PX & \rm J1846-7403 \\
-23.171 & -0.375 & 3.40 & 0.0 & 1.9 & 0.03 & 0 & MPX & \rm G336.829-0.375(HII) \\
-15.000 & 1.100 & 0.5 & 50.0 & 1.6 & 0.14 & 0 & MP & \rm Gum55(HII) \\
-9.000 & 24.000 & 2.00 & 0.010 & 0.136 & 0.011 & 1 & P & \rm J1614-2230(HII) \\
-3.459 & -7.766 & 2.65 & 0.0 & 11.102 & 0.01 & 0 & PX & \rm J1809-3547 \\
-1.660 & -1.850 & 0.780 & 0.0 & 1.5 & 0.068 & 0 & P & \rm Ter6(HII) \\
-0.056 & -0.046 & 10.0 & 400.0 & 2.400 & 0.0038 & 0 & P & \rm J1745-2900 \\
2.794 & -49.466 & 0.13 & 0.5 & 0.083 & 0.01 & 0 & P & \rm J2144-3933 \\
7.000 & 23.500 & 1.85 & 1.1 & 0.112 & 0.014 & 1 & MP & \rm J1623-0908(HII) \\
14.794 & 9.180 & 0.04 & 5.5 & 0.2 & 0.01 & 0 & P & \rm J1744-1134 \\
27.456 & 75.728 & 5.806 & 0.0 & 3.096 & 0.01 & 0 & PX & \rm J1354+2453 \\
27.657 & 2.272 & 1.586 & 1.2 & 1.61 & 0.053 & 0 & P & \rm J1833-0338 \\
31.303 & -0.134 & 21.348 & 0.0 & 2.9 & 0.01 & 0 & P & \rm Glimpse-C01(HII) \\
38.354 & 2.064 & 1.0 & 50.0 & 2.75 & 0.07 & 0 & P & \rm J1853+0545 \\
38.365 & 0.062 & 0.203 & 0.0 & 9.5 & 0.04 & 0 & PX & \rm J1901+0459(HII) \\
46.170 & 0.180 & 0.726 & 0.0 & 8.2 & 0.01 & 0 & PX & \rm J1915+1150 \\
49.070 & -0.380 & 2.1 & 200.0 & 5.00 & 0.090 & 0 & MPX & \rm W51-II(HII) \\
49.070 & -0.380 & 2.1 & 0.0 & 5.29 & 0.138 & 0 & MPX & \rm W51-I(HII) \\
49.791 & -61.252 & 1.306 & 0.0 & 3.422 & 0.01 & 0 & PX & \rm J2257-16 \\
51.979 & 0.542 & 1.5 & 0.0 & 5.29 & 0.03 & 0 & MPX & \rm G051.979+00.542(HII) \\
53.935 & 0.229 & 0.7 & 0.0 & 7.0 & 0.147 & 0 & MPX & \rm W52(HII) \\
56.750 & -4.560 & 6.06 & 0.0 & 2.0 & 0.01 & 0 & P & \rm M71 \\
58.966 & 1.814 & 0.746 & 0.0 & 8.7 & 0.01 & 0 & PX & \rm J1934+2352 \\
59.931 & -1.423 & 0.50 & 73.0 & 4.155 & 0.01 & 0 & P & \rm J1949+2306 \\
60.395 & -0.131 & 1.057 & 0.0 & 8.5 & 0.01 & 0 & PX & \rm J1945+2410\_P \\
63.598 & 1.687 & 1.50 & 25.0 & 2.37 & 0.05 & 0 & MP & \rm J1946+2611(HII) \\
68.564 & 1.329 & 1.26 & 0.0 & 7.1 & 0.01 & 0 & PX & \rm J1958+3156 \\
69.042 & -0.535 & 0.50 & 64.0 & 3.437 & 0.01 & 0 & MP & \rm J2007+3120 \\
77.000 & 1.100 & 0.9 & 4.7 & 1.5 & 0.179 & 1 & MPX & \rm CygnusI(HII) \\
78.400 & 1.200 & 6.0 & 1.8 & 1.65 & 0.032 & 1 & MPX & \rm CygnusII(HII) \\
79.133 & -5.589 & 2.366 & 0.075 & 1.5 & 0.01 & 0 & P & \rm J2055+3630 \\
85.325 & -0.800 & 1.3 & 10.0 & 1.5 & 0.094 & 1 & MPX & \rm S117(HII) \\
89.003 & -1.266 & 6.457 & 0.007 & 1.111 & 0.01 & 0 & P & \rm J2113+4644 \\
90.857 & 1.690 & 2.3 & 0.0 & 5.85 & 0.086 & 0 & MPX & \rm G090.856+01.69(HII) \\
104.255 & 7.412 & 4.227 & 0.011 & 1.389 & 0.01 & 0 & P & \rm J2149+6329 \\
108.860 & 6.150 & 4.0 & 0.045 & 0.730 & 0.01 & 0 & P & \rm J2225+6535(HII) \\
113.138 & 3.079 & 1.308 & 0.0 & 2.9 & 0.01 & 0 & PX & \rm J2319+6411 \\
121.298 & 0.659 & 2.00 & 0.0 & 0.929 & 0.025 & 0 & PX & \rm J0026+6320(HII) \\
126.398 & 69.947 & 3.009 & 0.0 & 3.194 & 0.01 & 0 & PX & \rm J1244-4708 \\
128.100 & 0.950 & 5.0 & 0.0 & 4.0 & 0.039 & 0 & MPX & \rm J0137+6349 \\
130.719 & 3.084 & 0.859 & 0.0 & 2.3 & 0.01 & 0 & PX & \rm J0205+6449 \\
136.880 & 0.912 & 13.0 & 0.0 & 1.95 & 0.015 & 0 & MPX & \rm G136.884+0.911(HII) \\
160.363 & 3.077 & 0.2 & 50.0 & 0.765 & 0.01 & 0 & P & \rm J0502+4654 \\
171.960 & -2.170 & 2.56 & 0.0 & 0.500 & 0.02 & 0 & PX & \rm J0517+3436\_P(HII) \\
\enddata
\tablecomments{Clumps are modeled as spherical Gaussians with properties listed left to right: Central Galactic coordinates, peak electron density, fluctuation parameter, distance, radius, edge ($0 =$ exponentially rolls off to $5r_c$, $1=$ truncated at the $1/e$ point), type ($\rm P=$ pulsar, $\rm MP$= multiple pulsars, $\rm X=$ pulsar with $\rm DM_{obs}>DM_{mod}(\infty)$), name (pulsar J-names are given; the cataloged \HII region name is given when multiple pulsars are behind it, and all clumps based on known \HII regions are noted in parentheses).}
\end{deluxetable*}

\startlongtable
\begin{deluxetable*}{C C C C C C C C C l L}\label{tab:voids}
\tabletypesize{\scriptsize}
\tablecaption{Voids New or Revised in \model}
\tablehead{\colhead{$l$ $(^{\circ})$} & \colhead{$b$ $(^{\circ})$} & \colhead{$n_{e}$ (cm$^{-3}$)} & \colhead{$F_v$} & \colhead{$D_v$ (kpc)} & \colhead{$aav$ (kpc)} & \colhead{$bbv$ (kpc)} & \colhead{$ccv$ (kpc)} & \colhead{Edge} & \colhead{Type} & \colhead{Name}}
\startdata
-108.30 & 40.515 & 0.001 & 0.00 & 0.50 & 0.23 & 0.01 & 0.01 & 1 & P & \rm J1024-0719\\
-54.100 & -44.890 & 0.002 & 0.00 & 1.00 & 0.50 & 0.05 & 0.05 & 1 & P & \rm 47Tuc \\
-22.98 & 13.27 & 0.001 & 0.00 & 1.20 & 0.20 & 0.30 & 0.10 & 1 & P & \rm NGC5986 \\
-22.98 & 13.27 & 0.001 & 0.00 & 4.00 & 0.30 & 0.60 & 0.20 & 1 & P & \rm NGC5986 \\
-16.901 & -2.683 & 0.090 & 0.10 & 1.00 & 0.40 & 0.10 & 0.10 & 1 & P & \rm J1709-4428\\
-7.33 & 19.45 & 0.001 & 0.00 & 2.00 & 0.20 & 0.50 & 0.20 & 1 & P & \rm M80 \\
-7.33 & 19.45 & 0.001 & 0.00 & 4.00 & 0.20 & 1.00 & 0.20 & 1 & P & \rm M80 \\
-0.269 & -19.596 & 0.002 & 5.00 & 0.70 & 0.05 & 0.20 & 0.10 & 1 & P & \rm J1909-3744\\
-0.056 & -0.046 & 10.00 & 0.00 & 8.50 & 0.013 & 0.145 & 0.013 & 0 & P & \rm J1745-2900 \\
0.000 & -10.00 & 0.001 & 0.00 & 8.60 & 4.0 & 4.0 & 1.0 & 1 & MP & \rm Fermi-South\\
0.000 & 10.00 & 0.001 & 0.00 & 8.60 & 4.0 & 4.0 & 1.0 & 1 & MP & \rm Fermi-North \\
2.212 & -10.517 & 0.005 & 0.00 & 1.50 & 0.30 & 0.80 & 0.30 & 1 & MP & \rm Car-Sgr \\
4.900 & 9.72 & 0.001 & 0.00 & 1.40 & 0.10 & 0.60 & 0.20 & 1 & P & \rm NGC6342 \\
4.900 & 9.72 & 0.001 & 0.00 & 3.50 & 0.10 & 0.60 & 0.20 & 1 & P & \rm NGC6342 \\
23.109 & 0.257 & 0.001 & 0.00 & 1.50 & 0.20 & 0.40 & 0.05 & 1 & P & \rm J1832-0836 \\
46.564 & 1.795 & 0.001 & 0.00 & 3.00 & 0.80 & 0.70 & 0.20 & 1 & P & \rm J1910+1256 \\
79.133 & -5.589 & 0.001 & 0.00 & 4.50 & 1.50 & 0.30 & 0.20 & 1 & P & \rm J2055+3630 \\
129.150 & 2.530 & 0.010 & 0.00 & 2.15 & 0.50 & 0.70 & 0.50 & 1 & MP & \rm PerseusI \\
144.995 & -1.221 & 0.001 & 0.00 & 1.40 & 0.20 & 0.40 & 0.10 & 1 & P & \rm J0332+5434\\
149.099 & -0.522 & 0.030 & 0.00 & 2.00 & 0.15 & 0.25 & 0.05 & 1 & P & \rm J0357+5236\\
163.909 & 18.643 & 0.002 & 60.0 & 0.80 & 0.34 & 0.01 & 0.01 & 1 & P & \rm J0636+5128\\
168.275 & -1.187 & 0.001 & 0.00 & 2.00 & 0.20 & 0.70 & 0.10 & 1 & P & \rm J0509+3801\\
188.800 & 4.100 & 0.001 & 0.00 & 2.10 & 0.20 & 0.31 & 0.30 & 1 & MP & \rm PerseusII \\
\enddata
\tablecomments{Voids are modeled as triaxial ellipsoidal Gaussians with properties listed left to right: Central Galactic coordinates, peak electron density, fluctuation parameter, distance, major and minor axes $(aav,bbv,ccv)$, edge (defined as for clumps in Table~\ref{tab:clumps}), type (same as Table~\ref{tab:clumps}) and name (voids constrained by multiple pulsars are named for their foreground structures -- either the Fermi Bubbles or the spiral arm in which the void is placed).}
\end{deluxetable*}

\bibliography{master_bib}

\end{document}